\documentclass[usenatbib]{mn2e}

\usepackage{natbib,epsfig,times}

\title[MUNICS -- IV. Biases in the Completeness of Near-Infrared
Imaging Data]{The Munich Near-Infrared Cluster Survey -- IV. Biases in the
  Completeness of Near-Infrared Imaging Data}

\author[J. Snigula et al.]  
{J.~Snigula,$^1$
  N.~Drory,$^1$
  R.~Bender,$^1$
  C.~S.~Botzler,$^1$ 
  G.~Feulner,$^1$
  U.~Hopp$^1$\\
  $^1$Universit\"ats-Sternwarte M\"unchen, Scheinerstr. 1, 
  D-81679 M\"unchen, Germany}
 
\date{Accepted --; Received --; }

\pagerange{\pageref{firstpage}--\pageref{lastpage}} \pubyear{2001}

\begin{document}

\label{firstpage}

\maketitle

%
%
\begin{abstract}
   We present the results of completeness simulations for the
  detection of point sources as well as redshifted elliptical and
  spiral galaxies in the $K'$-band images of the Munich Near-Infrared
  Cluster Survey (MUNICS).  The main focus of this work is to quantify
  the selection effects introduced by threshold-based object detection
  algorithms used in deep imaging surveys. Therefore, we simulate
  objects obeying the well-known scaling relations between effective
  radius and central surface brightness, both for de Vaucouleurs and
  exponential profiles. The results of these simulations, while
  presented for the MUNICS project, are applicable in a much wider
  context to deep optical and near-infrared selected samples. We
  investigate the detection probability as well as the reliability for
  recovering the true total magnitude with Kron-like (adaptive)
  aperture photometry.  The results are compared to the predictions of
  the visibility theory of Disney and Phillipps in terms of the
  detection rate and the lost-light fraction.  Additionally, the
  effects attributable to seeing are explored. The results show a bias
  against detecting high-redshifted massive elliptical galaxies in
  comparison to disk galaxies with exponential profiles, and that the
  measurements of the total magnitudes for intrinsically bright
  elliptical galaxies are systematically too faint. Disk galaxies, in
  contrast, show no significant offset in the magnitude measurement of
  luminous objects.  Finally we present an analytic formula to predict
  the completeness of point-sources using only basic image parameters.
  
\end{abstract}


%
%

\section{Introduction}
\label{s:introduction}

The luminosity function as well as the mass function, and to a lesser
degree the number counts of galaxies provide an important
observational toolset for understanding the evolution of galaxies.
Due to their statistical nature, these methods rely on the
understanding of sample selection effects, i.e.\ the knowledge of what
fraction of the true number of galaxies, as a function of their
intrinsic properties, is actually present in the sample. In order to
extract this information from the data, two different approaches are
commonly used in the literature, creation of artificial objects or
modification of observed objects.

The synthetic objects are, as described by e.g. \citet{Martini2001},
usually point-sources or objects with galactic profiles, that are
inserted into the observed images. Then the fraction of objects
recovered by the applied detection algorithm as a function of, e.g,
the assigned apparent magnitude is computed. This approach exhibits
two major drawbacks: The galaxies are created using a discrete set of
half-light radii and a continuous range of magnitudes, thus
ignoring known relations between surface-brightness and effective
radius, like the fundamental plane \citep{BBF92,GFSBP2000}. The
results show the completeness limits for distinct types of objects,
but as the actually observed ratio of these types is unknown,
predictions about the total completeness of the data cannot be made.
  
In the second case, as conducted by e.g.\citet{BLK1998},
\citet{ADCZFG99} or \citet{Saracco99}, images of the observed objects
are dimmed or brightened to produce artificial objects. The advantage
of this approach is, that it preserves the observed mix of galaxies, 
assuming that it remains unchanged with time, yet the resulting
completeness fractions for extended objects are questionable, as long
as the resolution of images is seeing-limited.  When the bright
extended objects are dimmed, they are still profile dominated, whereas
in reality a faint object would be seeing dominated. The same holds true
for artificially brightened faint sources. The resulting objects would
still be seeing dominated.  Furthermore, the observed size of a local
object of given magnitude is different from the size of a distant
object of the same brightness.  Consequently this method only yields
information about the probability to detect an object if it was
fainter, but not about the probability to detect an intrinsically
faint object.

Another drawback of both approaches is that the obtained results can
only be used to define the completeness of the survey in apparent
magnitudes. The magnitudes and radii of the artificially constructed
objects all are physically plausible, but the objects do not occupy
one plane in the $\langle \mu_e \rangle - r_e - z$ parameter space.
Therefore the results derived using galactic profiles are not usable
to correct absolute magnitudes, as effects of the completeness of
different galaxy types at given redshifts are unknown.

In deep extra-galactic surveys, the observed galaxies span a wide
range in intrinsic profile shape, intrinsic brightness, and intrinsic
size. The apparent quantities vary with cosmological distance, such
that the fraction of galaxies visible is also a function of redshift.
The need to simulate objects obeying the known scaling relations for
galaxies was pointed out several times in the literature.  Using
profiles obeying the magnitude-radius relations,~\citet{Yoshii1993}
predicted the number of objects lost in number count analysis. He
finds a strong dependence of the detection rate on the applied
detection criteria and magnitude measurement algorithm, leading to a
larger number of undetected faint galaxies at high redshifts. 

\cite{Dalcanton1998} analysed the biases of the luminosity function
introduced by the cosmological effects of size variation with
redshift, and cosmological dimming for galactic profiles in dependence
of size and magnitude, taking into account effects of seeing. Starting
from the deficiency that magnitudes are usually measured as some sort
of isophotal magnitude, that is directly influenced by the above
mentioned effects, she finds the possibility of a severe
underestimation of the true luminosity function introduced by the fact
that an object and distance dependent part of the light is lost
outside the limiting isophote.

The simulations presented in this paper were performed by adding
artificially created objects into the $K'$-band images of the Munich
Near-Infrared Cluster Survey (MUNICS; \citealp{MUNICS1}). We analysed
the detection probability and the reliability for recovering the
assigned magnitude for point-like sources (Moffat profiles),
elliptical galaxies (de Vaucouleurs profiles), and spiral galaxies
(exponential profiles).  The radii and magnitudes of the simulated
galaxies were distributed according to the projected fundamental-plane
relation~\citep{BBF92} for ellipticals and a Freeman law type
relation~\citep{Freeman1970} constructed from observed data for
spirals.  The galaxies were simulated at five distinct redshifts
between $z=0.5$ and $z=1.5$, taking into account size variation with
redshift as well as cosmological dimming and K-correction. A flat
universe with $\Omega_M = 0.3$, $\Omega_\Lambda=0.7$ and
$H_0=65$~km~s$^{-1}$~Mpc$^{-1}$ was assumed.

The results of the completeness simulations for the different fields
of the MUNICS survey presented here, illustrate the principal
limitations of imaging surveys. A full correction of the
incompleteness would only be possible if the type mix of galaxies was
known.  The presentation of the results here is limited to the
$K'$-band data, but as the selection biases are caused by the physical
nature of the objects, they are applicable in a much wider context to
deep extragalactic surveys spanning the optical and near-infrared
wavelength regimes. The conclusions drawn here result from the use of
a threshold-based detection algorithm.  Accordingly they will hold
true for other similarly created datasets as well.

Finally, a reliable and handy analytic formula to estimate the
completeness limit of a survey for point-like sources is presented.
In Section~\ref{s:implementation} the implementation of the
simulations and the generation of the artificial objects is described.
Section~\ref{s:discussion} presents the results of the simulations,
and a discussion of these. To compare the results with analytic
predictions, we present the results of an analysis using the
visibility theory devised by \citet{DP1983} in
Section~\ref{s:visibility} for the detection probability and in
Section~\ref{s:lostlight} for the lost-light fraction. In
Section~\ref{s:seeing} we discuss the effects of seeing in the above
mentioned analysis. Finally, in Section~\ref{sec:values} we present
the results of the completeness simulations for the MUNICS fields.

\subsection{The Munich Near-Infrared Cluster Survey (MUNICS)}

The Munich Near-Infrared Cluster Survey (MUNICS) is a wide-field
medium-deep survey in the near-infrared $K'$ and $J$ pass-bands (Drory
et al.\ 2001; MUNICS1 hereafter). The survey consists of a
$K'$-selected catalogue down to $K' \le 19.5$ covering an area of 1
square degree. Additionally, 0.35 square degrees have been observed in
$I$, $R$, and $V$. The layout of the survey, the observations and data
reduction are described in \citet{MUNICS1}.

\section{Implementation of the simulations}
\label{s:implementation}

Detection probabilities and photometry results were analysed for three
different profile shapes: de Vaucouleurs profiles, exponential disks
and point-like sources.

For each profile type and each image, 200 artificially created objects
were added to the image, taking into account the noise properties of
the background and the photometric zero-point. The resulting image was
processed in the same way as described in MUNICS1 regarding detection
and photometry, using the software package YODA~\citep{YODA01}. The
resulting object catalogue was used to calculate the fraction of
recovered objects and to compare the resulting photometry in Kron-like
elliptical apertures to the quantities in the input catalogue.

This procedure was repeated 500 times resulting in a total of 100 000
artificial objects per type. In each run, the artificial objects were
distributed randomly in (x,y)-position across the image, excluding a
25 pixel wide strip along the image borders, and requiring 20 pixel
distance to existing objects, as we intentionally avoid crowding which
would introduce effects beyond the scope of this work.  Extended
sources were convolved with a Gaussian with a FWHM of the measured
seeing in the image. Poisson noise was added to all profiles. In the
following we discuss the generation of objects of each of the examined
profile types in more detail.

\subsection{Stars and point-sources}
\label{s:stars}

Point-like sources were simulated using a Moffat
profile~\citep{Moffat1969} of the form
\begin{equation}
  I(r) = \frac{I_c}{\left( 1 + \left(\frac{r}{R}\right)^2 \right)^{\beta}}.
  \label{e:star_prof}
\end{equation}
with the characteristic intensity $I_c$. The size parameter $R$,
defining the radius of the created object, was set to the measured
value from point-like real sources in the analysed image, and the
Moffat parameter $\beta$ was fixed at the canonical value of $2.5$.
The apparent magnitudes assigned to the point-sources were chosen
randomly from a constant probability density in the range $15 \le
m_{K'} \le 25$.

\subsection{Elliptical galaxies}
\label{s:ellipticals}

Elliptical galaxies were simulated using a de Vaucouleurs profile
\citep{Vauco48}. For a galaxy with effective radius $r_e$ and
effective intensity $I_e$ (defined as the intensity at $r_e$), this
profile can be written as
\begin{equation}
  \label{e:devauc}
    I(r) = I_e \exp \left( -7.67 
      \left( \frac{r}{r_e} \right)^{\frac{1}{4}} \right).
\end{equation}

For the sake of simplicity an $r^{1/4}$ law is applied, ignoring
relations between luminosity and the shape of the profile. But it
should be kept in mind, that an $r^{1/n}$ profile with $n>4$ would
show the same biases as described here. To create a realistic
population of elliptical galaxies, the radii and absolute magnitudes
of the galaxies were distributed according to the Kormendy relation
\citep{Kormendy1985}, a projection of the local fundamental-plane
relation \citep{BBF1997}.  We used the $K$-band Kormendy relation
published by \citet{PDC1995} for the effective radius of the galaxy
$R_e$ in kpc and the mean effective surface brightness $\langle \mu_e
\rangle$ in mag arcsec$^{-2}$

\begin{equation}
  \label{e:kormendy}
  \log R_e = 0.332 \langle \mu_e \rangle - 5.090 .
\end{equation}
Using the definition of $\langle \mu_e \rangle$
\begin{equation}
  \label{e:mag_mu}
    m = -1.995 - 5 \log r_e + \langle \mu_e  \rangle .
\end{equation}
equation~(\ref{e:kormendy}) can be transformed into a relation between
absolute magnitude $M$ and effective radius $R_e$
\begin{equation}
  \label{e:devauc_rel}
  M_{K'} = - 1.99 \log R_e -23.235.
\end{equation}
The absolute magnitudes were chosen randomly and uniformly in the
range $-26.5 \le M_{K'} \le -20.5$, corresponding to a range of $\pm
3$ magnitudes around $M_K^*$ for the local K-band luminosity function
\citep{Loveday00}.

To simulate a galaxy at redshift $z$, the apparent radius was
calculated from the given physical radius using the angular distance,
and the absolute magnitude was transformed into the apparent magnitude
using the luminosity distance. During each repetition of the
simulation 40 galaxies for each of the five redshifts $z \in \lbrace
0.5, 0.75, 1.0, 1.25, 1.5 \rbrace$ were created. The axis ratios $e$
were selected randomly in the range $0.7 \le e \le 1.0$, the position
angles were chosen arbitrarily. For the galaxies at $z=0.5$ the
absolute magnitude range was shifted $-25.5 \le M_{K'} \le -19.5$ to
better trace the faint-end drop-off of the completeness curve. The
resulting apparent magnitudes were adjusted using only K-corrections
derived from model SEDs from~\citet{photred00_mod}. The used values are
listed in Table~\ref{tab:kcorr}. Possible effects introduced by
luminosity evolution with redshift were explored, and are discussed at
the end of section~\ref{s:discussion}.

\begin{table}
  \centering
  \begin{tabular}{ccc}
    $z$ & de Vaucouleurs & exponential\\\hline
    $0.5$ & $-0.63$ & $-0.68$ \\
    $0.75$ & $-0.73$ & $-0.83$ \\
    $1.0$ & $-0.84$ & $-0.97$ \\
    $1.25$ & $-0.88$ & $-1.03$ \\
    $1.5$ & $-0.87$ & $-1.02$ \\
  \end{tabular}
  \caption{K-corrections in magnitudes for de Vaucouleurs and
    exponential profiles applied in the simulations. The values were
    derived from empirical SEDs~\citep{photred00_mod}.} 
  \label{tab:kcorr}
\end{table}

\subsection{Spiral galaxies}
\label{s:spirals}

\begin{figure}
  \centering
  \includegraphics[width=8.4cm]{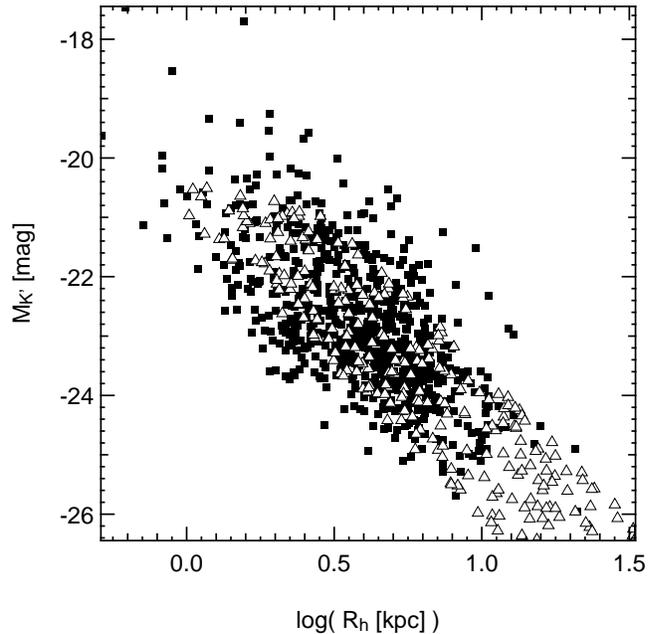}
  \caption{Absolute $K'$-band Magnitudes $M_K'$ and effective radii for 
    the galaxies fitted by a pure exponential profile published by
    \citet{GFSBP2000} (filled squares) and some during the simulations
    created artificial objects (open triangles).}
  \label{f:spiral_dist}
\end{figure}

 Spiral galaxies were simulated as pure exponential profiles
with no bulge component. An analysis of the bulge-disk decomposition
for the galaxies published by \citet{GFSBP2000} shows that the
contribution of the bulge to the total luminosity is less than 18\%.
We therefore neglect the contribution of the bulge. 

For a galaxy with half-light radius $r_h$ and central intensity $I_e$
the intensity profile can be written as
\begin{equation}
  \label{e:exponential}
  I(r) = I_0 \exp \left( -1.6783 \frac{r}{r_h} \right).
\end{equation}

To create objects with realistic magnitude-size ratios, all galaxies
fitted by a pure exponential profile from $H$-band surface photometry
data of spiral galaxies published by \citet{GFSBP2000} were used. The
absolute $H$-band magnitudes were transformed into the $K'$-band using
$H-K'$ colours derived from empirical SEDs presented
in~\citet{photred00_mod}. The mean correction was $H-K' \simeq 0.128$ mag.
Using the objects' distances, the half-light radii in kpc $R_H$ were
computed from the published values. The resulting distribution of
galaxies in the $\log R_h - M_{K'}$ plane was then approximated by
assuming
\begin{equation}
  \label{e:spirals}
  M_{K'} = -5 \left( \log R_h \pm 0.25 \right) - 19.7. 
\end{equation} 
For a population of local spiral galaxies, this distribution would
be analogous to a Freeman law with a $K'$-band central surface
brightness of $17.52$ mag arcsec$^{-2}$ and an rms of $1.73$ mag.

Fig.~\ref{f:spiral_dist} shows the absolute magnitudes and effective
radii of the Gavazzi et al.\ data and of some artificially created
objects. The distribution of the artificial objects reproduces that
of the observed population reasonably well, extending to somewhat
brighter magnitudes.

The absolute magnitudes of the objects were randomly chosen in the
range $-26.5 \le M_{K'} \le -20.5$, again corresponding to a range of
$\pm 3$ magnitudes around $M^*_K$ for a local K-band luminosity
function~\citep{Loveday00}. Using the same techniques as for the
elliptical galaxies, the spirals were simulated for the same five
redshifts, with axis ratios in the range $0 < e \le 1$ and arbitrary
position angles. The resulting apparent magnitudes were adjusted using
only K-corrections derived from model SEDs shown in~\citet{photred00_mod},
no evolution corrections were applied. The results of tests for
influences of luminosity evolution are discussed at the end of
section~\ref{s:discussion}. The applied values are listed in
Table~\ref{tab:kcorr}.  The mean surface brightnesses $\langle \mu
\rangle$ were derived from a sample of local spiral galaxies.
Brightness evolution with redshift would make the galaxies brighter,
while the radii are known not to change significantly up to redshifts
of $z \sim 1$.

\section{Results and Discussion}
\label{s:discussion}

\begin{figure}
  \centering
  \includegraphics[width=8.4cm]{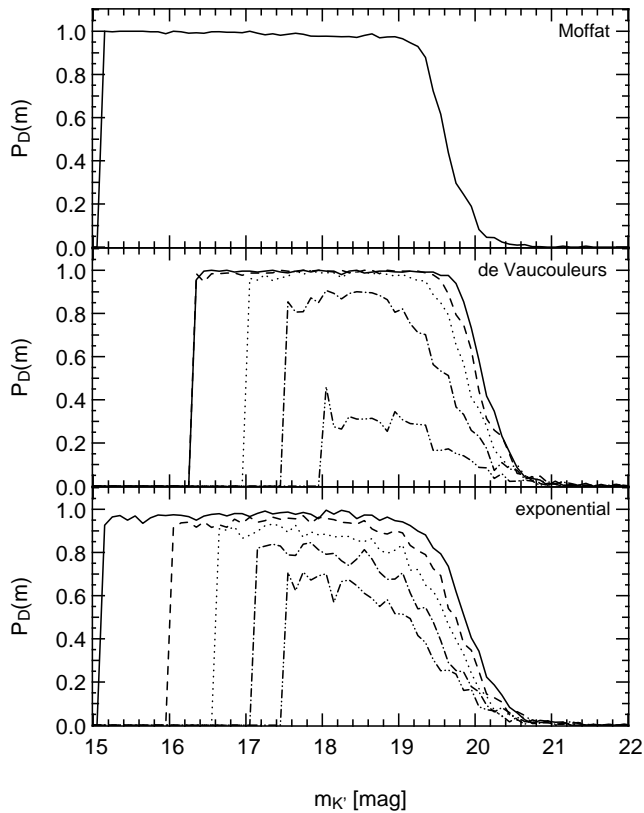}
  \caption{Probability $P_D(m)$ to re-detect an artificially created 
    point-source (upper panel), an extended object with a de
    Vaucouleurs profile (middle panel) or exponential profile (lower
    panel) as a function of the objects' $K'$-band input magnitude
    $m_{K'}$. The different line types in the middle and lower panel
    show the completeness for the five analysed redshifts: $z=0.5$
    (solid), $0.75$ (dash), $1.0$ (dot), $1.25$ (dashdot) and $1.5$
    (dashdotdot).}
  \label{f:compl_results}
\end{figure}

\begin{figure}
  \centering
  \includegraphics[width=8.4cm]{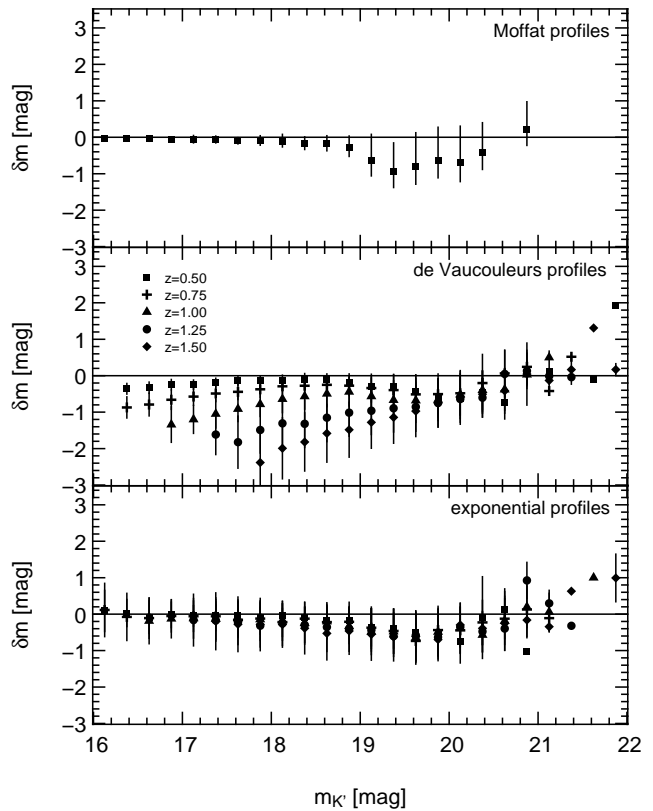}
  \caption{Mean magnitude difference $\delta m = m_{K'} - m_{meas}$
    between the assigned magnitude $m_{K'}$ of an object and the
    Kron-magnitude measured by the photometry software $m_{meas}$ in
    bins of $0.25$ mag as a function of the input $K'$-band magnitude
    $m_{K'}$. The top panel shows the results for point-sources, the
    middle panel for de Vaucouleurs profiles and the lower panel for
    exponential profiles. The errorbars show the standard deviation of
    $\delta m$}
  \label{f:compl_magmag}
\end{figure}

The results for one of the MUNICS fields with a seeing of
approximately one arcsec FWHM are shown in Fig.~\ref{f:compl_results}.
The upper panel shows the results for the point-sources, the middle
and the lower panel for the de Vaucouleurs and exponential profiles,
respectively, for the five redshifts simulated. An interesting effect
seen in the figure is that, for higher redshifts $z \ga 1$, the
detection probability does not reach unity even for the brightest
objects, forming a plateau at some lower value. This effect is mainly
caused by the cosmological surface brightness dimming and discussed in
detail in Section~\ref{s:discussion} below.
Fig.~\ref{f:compl_results} shows that the completeness for point-like
sources provides a rough but reasonable approximation for the
completeness of the analysed extended objects up to a redshift of
$z=1.0$.

For each detected object, the difference between the input magnitude
and the measured Kron-magnitude was computed.
Fig.~\ref{f:compl_magmag} shows the mean magnitude differences for
the analysed profiles, averaged in bins of 0.25~mag and with the
standard deviation of the measurements in the bin indicated as
errorbars.

The object recovery probabilities for the high-redshift de Vaucouleurs
profiles, and to a lesser degree for the exponential profiles exhibit
a significant detection bias compared to lower redshift objects, as
shown in Fig.~\ref{f:compl_results}. The fact that the objects at
these redshifts never reach a detection probability of one is caused
by the distribution of their physical parameters along the
fundamental-plane relation -- in case of the ellipticals -- and
according to the Freeman law -- in case of the disks.  In both cases,
as the object's size increases its radius grows and therefore the
average surface brightness decreases. As a result, even the brightest
objects of the distribution fail to produce a large enough area above
the threshold in surface brightness to be detected with high
probability in the presence of noise. In Section~\ref{s:visibility}
these results are compared with the theoretical predictions of the
visibility theory, which will confirm the above statement.

The detection deficiencies at low redshifts found for the exponential
profiles are caused by the flat light distribution of the
objects, resulting in a low central surface brightness even for the
luminous objects. The image noise can then scatter these objects below
the detection threshold. This result is in general agreement with the
predictions of~\citet{GBS1995}. In their paper they re-derive the
formalism devised by~\citet{DP1983} using two parameters for the disk
galaxies. They predict that disk galaxies would suffer from detection
biases introduced by their low central surface brightness. 

\citet{Martini2001} performed simulations similar to the ones
presented here, analysing the detection probabilities of Moffat, de
Vaucouleurs and exponential profiles. The galactic profiles were
created for a discrete set of half-light radii and a continuous range
of apparent magnitudes. The simulations show a decrease of the
limiting magnitude with increasing half-light radius. This effect can
be found in the simulations presented here, by comparing the
completeness magnitude at different redshifts. With increasing
redshift, the sampled range in absolute magnitudes is shifted toward
fainter apparent magnitudes, resulting in an increase of effective
radius at given apparent magnitude. The detection bias against high
redshifted elliptical galaxies we showed cannot be found by the
simulations performed by Martini, as these effects are only visible
when distributing the objects along a fundamental-plane relation.

The bias first shown by~\citet{DP1983}, that with increased distance
more low-luminosity system would be lost from the survey, is
reproduced by our simulations as well. With increasing redshift, the
magnitude of the completeness limit stays approximately constant, or
gets brighter, while the range of sampled absolute magnitudes is
shifted to fainter apparent magnitudes. This means, that with
increasing redshift, the observable range in absolute magnitudes moves
toward higher luminosities, the low-luminosity systems become
unobservable.

\begin{figure}
  \centering
  \includegraphics[width=8.4cm]{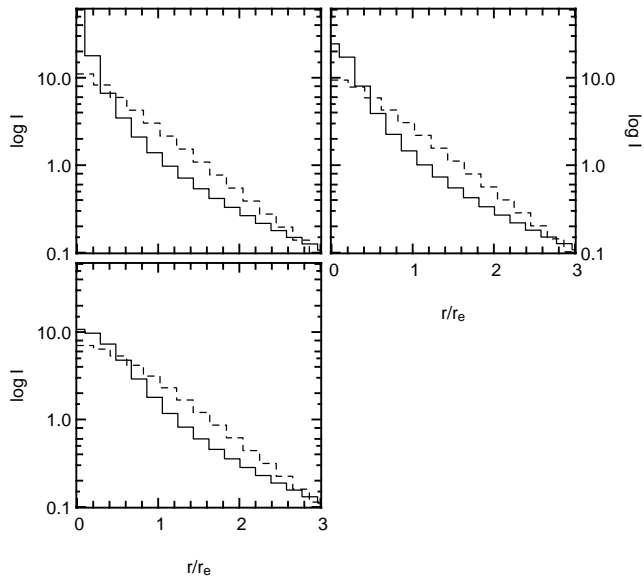}
  \caption{Intensity integrated over image pixels for a de Vaucouleurs
    profile (solid line) and an exponential profile (dashed line) at
    redshift $z=0.5$ with similar total magnitude and effective
    radius. The upper left panel shows the case in the absence of
    seeing, the upper right one, with a seeing of $0.5$ arcsec and the
    lower left with a seeing of $1.6$ arcsec. Plotted is the logarithm
    of the intensity against the radius relative to the effective
    radius.}
  \label{f:prof}
\end{figure}

The deviations of the measured magnitude from the true magnitude for
the de Vaucouleurs profiles as seen in Fig.~\ref{f:compl_magmag}, can
be explained as resulting from the estimate of the Kron-like aperture
radius under the conditions of a surface brightness limited detection
procedure and the involved intrinsic brightness profile of the
objects. The upper left panel of Fig.~\ref{f:prof} shows the intensity
of a de Vaucouleurs and an exponential profile with similar total
magnitude and effective radius, as a function of the radius in units
of the effective radius in absence of seeing. Assuming that both
profiles are detected at a similar intensity level, the Kron-radius of
the de Vaucouleurs profile would be smaller compared to the
exponential profile, even though the effective radius of both objects
is the same. The measured light within the underestimated radius
results then in a too faint object magnitude. Combined with the
previously mentioned surface brightness distribution along the
fundamental plane, this leads to an increased amount of lost flux and
a larger error in the output magnitude for brighter objects, as these
have lower surface brightnesses. In Section~\ref{s:lostlight} we will
confirm this explanation with predictions based on computations of the
visibility function and the lost-light fraction.

Our results confirm the predictions of~\citet{McGaugh1994},
that the magnitudes of distant spiral galaxies would be measured
correctly for luminous galaxies, and slightly underestimate them for
the systems with low-luminosity.

The deviations of the measured from the input magnitude we find, are
compatible with the simulations of \citet{Martini2001}, who compared
the reliability of different photometric methods for point-sources and
exponential profiles. For Kron-like magnitude measurements both shapes
would be measured correctly at the bright end, and slightly
underestimated at the faint end.

\begin{figure}
  \centering
  \includegraphics[width=8.4cm]{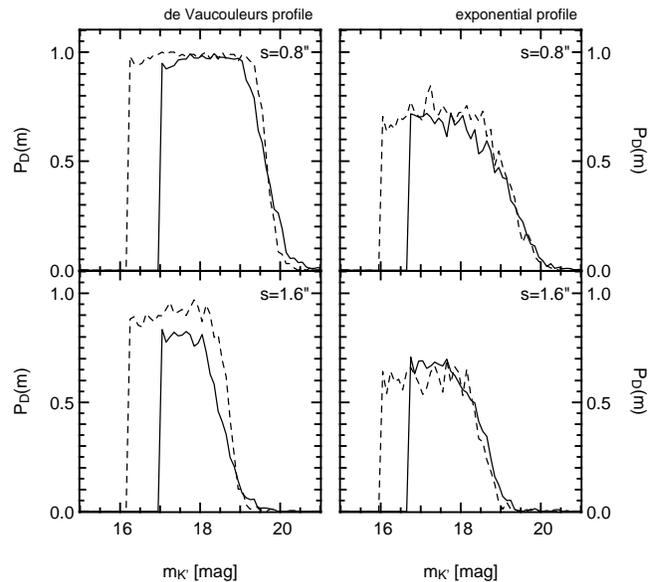}
  \caption{Results of a simulation assuming luminosity evolution with 
    redshift. Shown are the simulation results for de Vaucouleurs
    profiles (left column) and exponential profiles (right column) at
    a redshift $z=1$ for a seeing of $0.8"$ (upper row) and $1.6"$
    (lower row). The solid curve shows the results obtained from the
    normal simulations, the dashed line for a simulation with objects
    brightened artificially by $0.75$ mag.}
  \label{f:lumevol}
\end{figure}

To explore the effects caused by the increase of the objects' surface
brightness introduced by luminosity evolution with redshift, an
additional set of simulations for objects at $z=1$ was created.  The
brightnesses of the created artificial de Vaucouleurs and exponential
profiles was increased by $0.75$ magnitudes~\citep{PSEDE99}.  The results shown in
Fig.~\ref{f:lumevol} show no significant change of the magnitude of
the 50\% completeness limit. The reason for this is that the increase
of the detectable area is small in the magnitude range of the 50\%
limit. The increase in size is larger for brighter magnitudes, but the
positive effect is mainly lost due to the already high detection
probability, or may result in a slightly higher plateau level.

\section{Visibility theory}
\label{s:visibility}

\begin{figure}
  \centering
  \includegraphics[width=8.4cm]{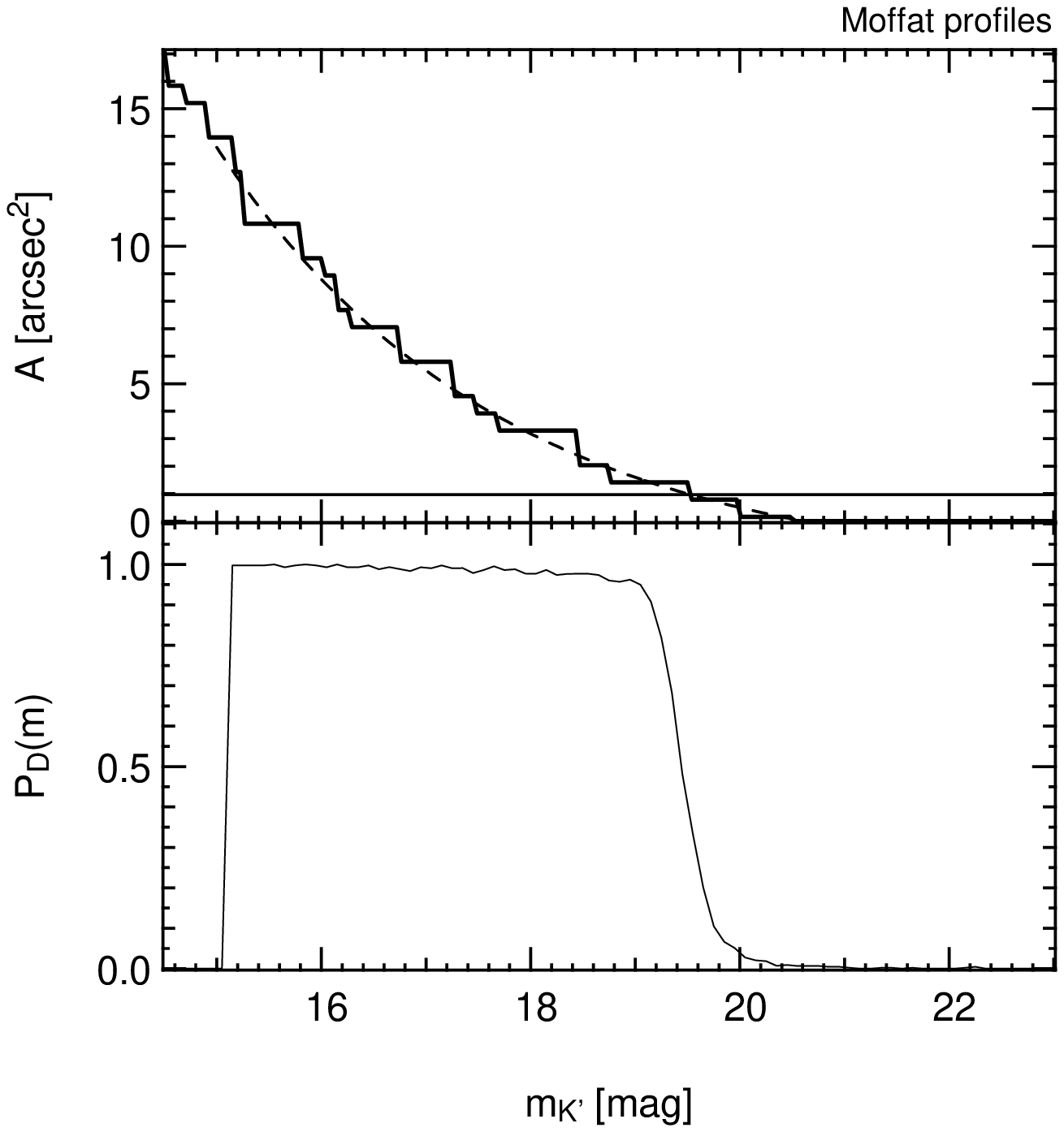}
  \caption{Area above detection threshold for point-sources with Moffat 
    profiles. The upper panel shows the predictions derived from the
    visibility theory~\citep{DP1983}. The thick solid histogram shows the
    area above the detection threshold integrated over the image
    pixels in presence of seeing.  The smooth dashed line gives the
    area above the surface brightness limit as predicted by the
    inversion of the profile.  The thin horizontal line is the area of
    the minimum number of connected pixels needed to detect an object.
    The lower panel shows the object recovery probability as a
    function of assigned input magnitude.}
  \label{fig:visfun-pointsource}
\end{figure}

\begin{figure*}
  \begin{minipage}{\textwidth}
  \includegraphics[width=8.4cm]{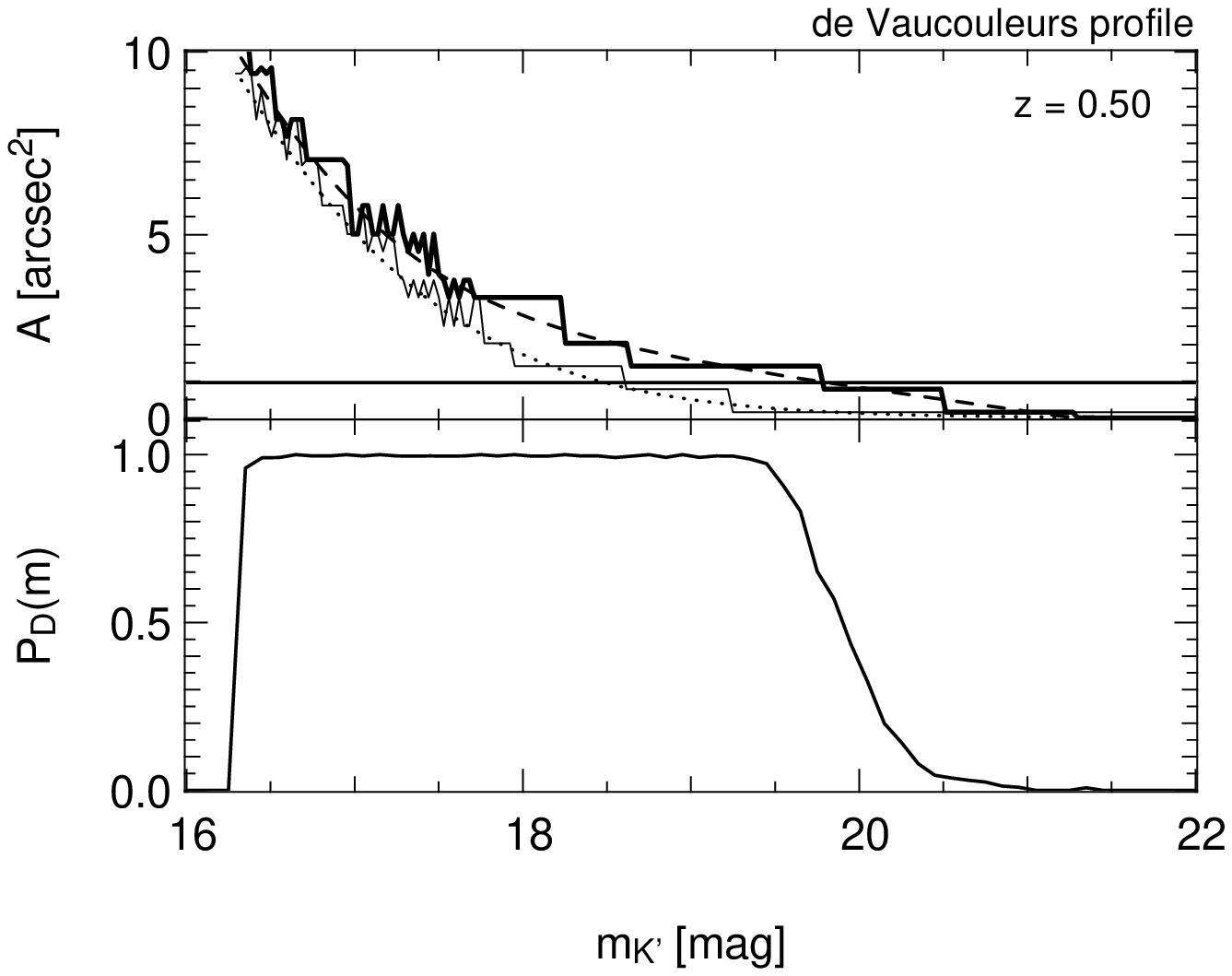}
  \includegraphics[width=8.4cm]{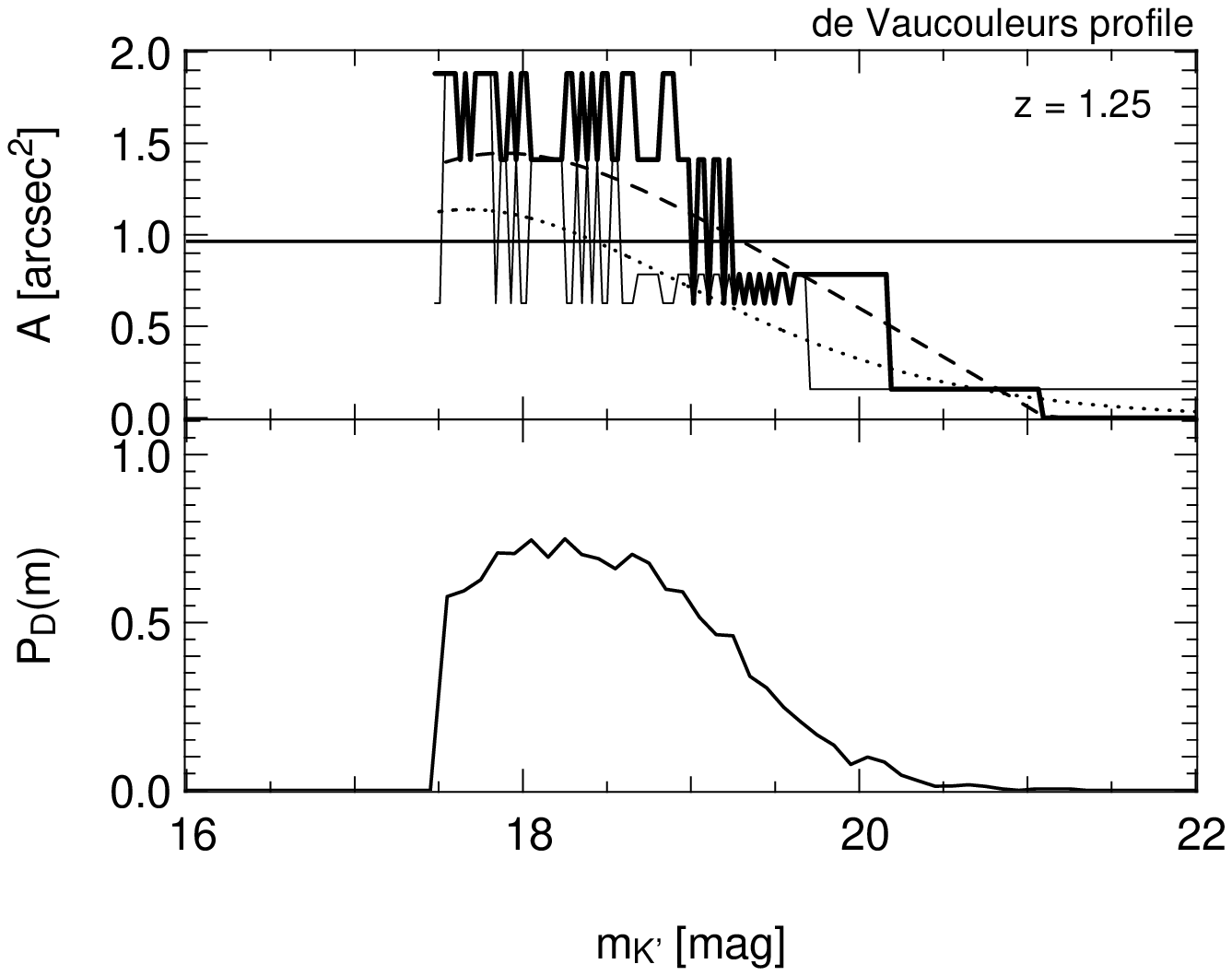}\vspace*{.5cm}
  \includegraphics[width=8.4cm]{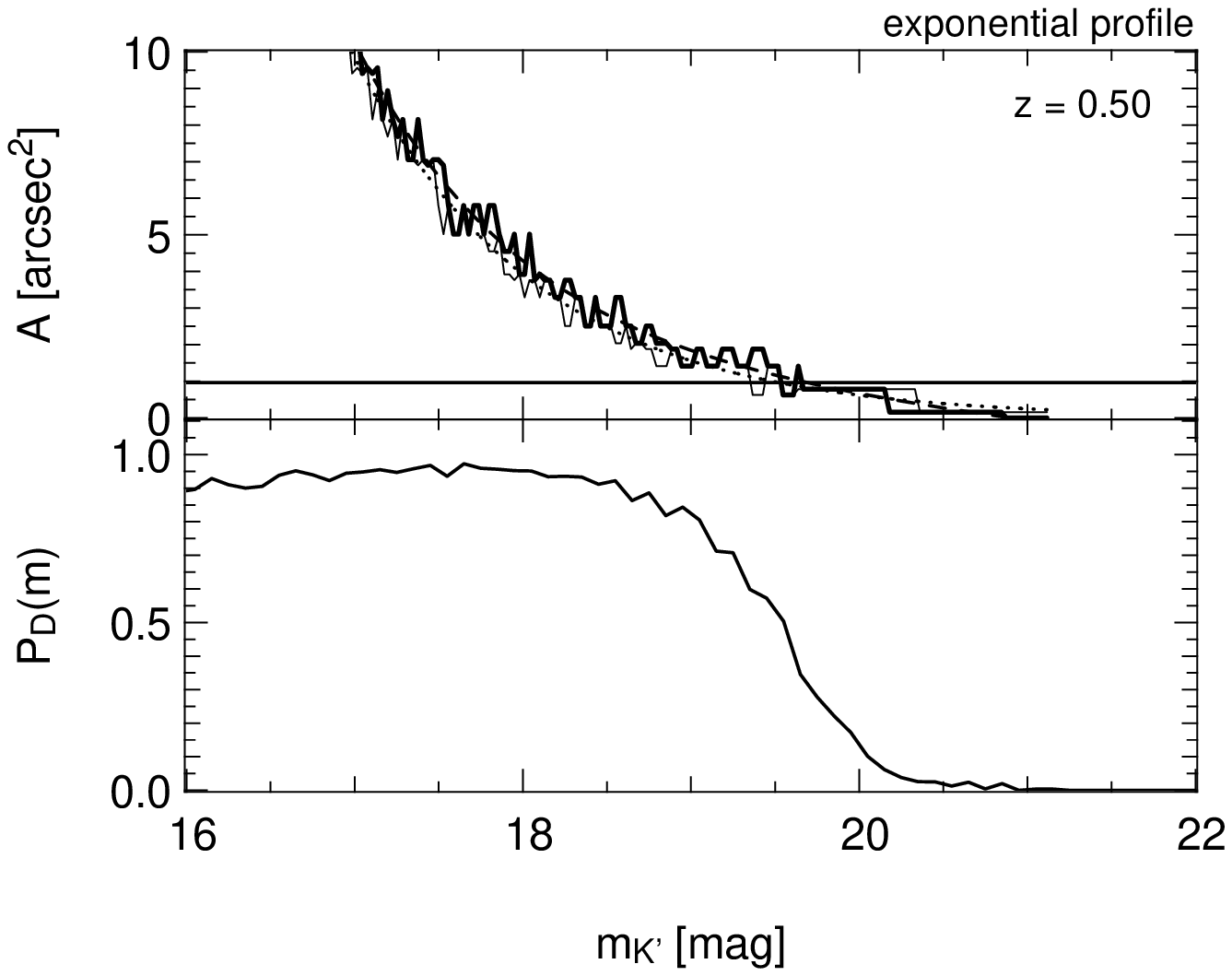}\hfill
  \includegraphics[width=8.4cm]{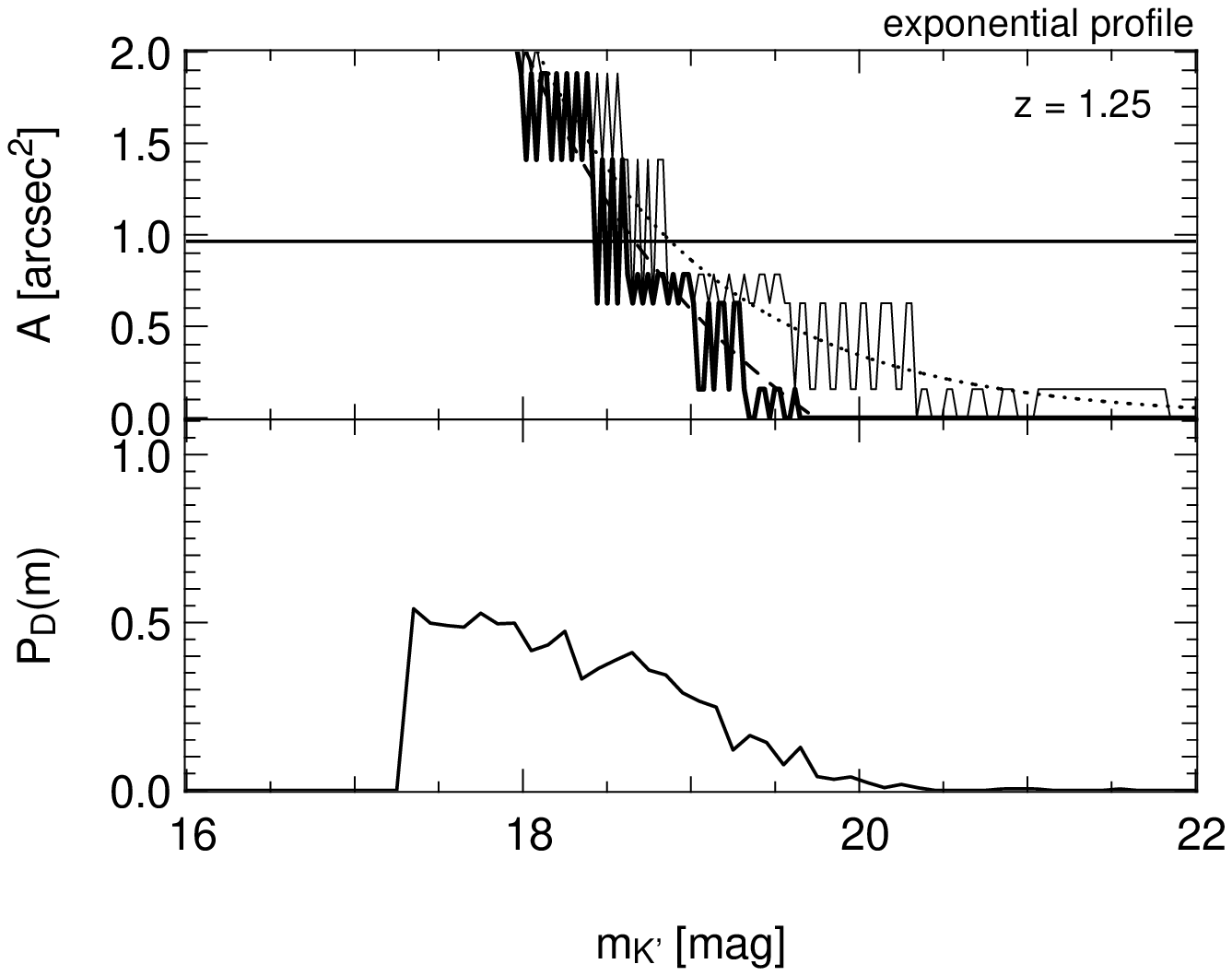}
  \caption{Predictions of the visibility function in comparison with
    the results of completeness simulations for de Vaucouleurs
    profiles (upper row) and exponential profiles (lower row) at
    redshift $z=0.5$ (left column) and $z=1.25$ (right column). The
    upper panel of each plot depicts the area above detection
    threshold. The thin solid line shows the area above the detection
    threshold integrated over the image pixels in absence of seeing,
    the thick solid line the same but in presence of appropriate
    seeing. The smooth dotted curve gives the area above the surface
    brightness limit in absence of seeing, the dashed line in presence
    of seeing. The thin horizontal line is the area of the minimum
    number of connected pixels needed to detect an object. The lower
    panels show the object recovery probability as a function of
    assigned input magnitude measured for a MUNICS image with seeing
    of $\sim$ 1 arcsec.}
  \label{f:area-compl}
  \end{minipage}
\end{figure*}

\citet{DP1983} and \citet{PDD1990} analysed the dependence of the
visibility of a galaxy on its surface brightness and apparent radius
in a survey with given detection constraints in surface brightness and
object radius. They estimated the maximum distance at which a galaxy
with a given magnitude and effective radius can be seen, by
calculating the distance at which the surface brightness at the
limiting radius drops below the detection threshold. 

Even though our goal is not to predict the maximum distance out to
which a given galaxy type can be seen, but to analyse biases that
occur when observing galaxies at high redshifts, this theoretical
approach can be used here as well to predict the behaviour of the used
threshold-based detection algorithms.

To detect an object we require a minimum number of consecutive pixels
to lie above a given brightness threshold, usually expressed in units
of the standard deviation of the background noise. The area required
to be above the threshold is usually determined by the seeing disk
(resolution element) size. Both values are adjusted such that faint
real objects are detected at a tolerable rate of false detections.

In our case the minimum number $N$ of consecutive pixels is chosen to
be $N = \pi ( 1.4 s/2 )^2$, with $s$ being the seeing FWHM in
the image. The threshold $t$ is set to three times the standard deviation
$\sigma$ of the local background noise, for details see~\citet{MUNICS1}.

In the case of circularly symmetric profiles, the calculation of the
area above $t$ is trivial, and the limiting surface brightness
$\mu_{lim}$ may be written as
\begin{equation}
  \label{e:mulim}
  \mu_{lim} = m_{zp} - 2.5 \log \frac{t}{p^2}
\end{equation}
for the magnitude zero-point $m_{zp}$ and the pixel scale $p$ (0.396
arcsec/pixel in MUNICS).

To create comparable completeness curves from the simulations as
discussed above, a set of simulations for point-sources with Moffat
profiles and circular face-on galaxies were calculated.  These results
are shown in Fig.~\ref{fig:visfun-pointsource} for point-sources and in
Fig.~\ref{f:area-compl} -- in the lower panel of each plot -- in the
upper row for de~Vaucouleurs profiles, and in the lower row for
exponential profiles at redshifts $z=0.5$ and $z =1.25$.

For Moffat profiles, the radius $R$ of the circular area $A$ above the
limiting isophote can be calculated as
\begin{equation}
  \label{e:r_moffat}
  R = s \sqrt{ 10^{\frac{\mu_{lim} - \mu_c}{2.5\beta}} - 1}
\end{equation}
with seeing $s$ and characteristic surface brightness $\mu_c$.  In the
upper panel of Fig.~\ref{fig:visfun-pointsource} the dashed line
shows the area $A$ resulting from the analytic calculation following
equation~(\ref{e:r_moffat}), the thick solid histogram gives the area integrated
over the image pixels for a simulated object. The thin horizontal line
indicates the minimum limiting area needed to detect an object.

Equation~(\ref{e:r_moffat}) can be transformed into
\begin{equation}
  \label{eq:get_compl}
  m = \mu_{lim} -2.5 \log \left( \frac{\pi s^2 \left( 1 + \left( \frac{R}{s}
  \right)^2 \right)^\beta}{\beta - 1} \right).
\end{equation}
Using equation~(\ref{e:mulim}) this can be written as
\begin{equation}
  \label{eq:get_compl2}
  m = m_{zp} -2.5 \log \left( \frac{t \pi s^2 \left( 1 + \left( \frac{R}{s}
  \right)^2 \right)^\beta}{p^2 \left( \beta - 1 \right)} \right).
\end{equation}
These formulas provide a simple way to estimate the completeness limit
for point-like sources for a given image, using easily measurable
parameters. Tests using the MUNICS data have shown that this formula
provides a robust estimate of the $\sim 50\%$ completeness level.

For de Vaucouleurs profiles with effective radius $r_e$ and effective
surface brightness $\mu_e$ within $r_e$, $R$ can be written as
\begin{equation}
  \label{e:r_devauc}
  R = r_e \left( \frac{\mu_{lim} - \mu_e}{8.3268} + 1 \right)^4
\end{equation}
and as 
\begin{equation}
  \label{e:r_expdisk}
  R = r_h \left( \frac{\mu_{lim} - \mu_0}{1.822} \right)
\end{equation}
for exponential profiles with half-light radius $r_h$ and the central
surface brightness $\mu_0$.

The dotted lines in the upper panels of Fig.~\ref{f:area-compl} show
the area within the limiting isophote as a function of apparent
magnitude, as calculated using equation~(\ref{e:r_devauc}) and
Equation~(\ref{e:r_expdisk}) for de Vaucouleurs profiles and
exponential profiles, respectively. For comparison, the same values
extracted from objects in the simulations are plotted as solid lines.
The area corresponding to the minimum area required for detection is
indicated as a horizontal line. The intersection of this line with the
curve then provides an estimate of the completeness limit.

The lower panels of these figures show the detection probability as a
function of apparent magnitude. Results are shown both with and
without seeing.  Seeing was modelled as a convolution with a Gaussian
profile. In the case of the de Vaucouleurs profiles, moderate seeing
($\la$ 1 arcsec) improves detectability, since it distributes flux
outwards, converting the steep core of the de Vaucouleurs profile to a
larger and flatter flux distribution. This effect is much weaker for
disk profiles since these are flatter, anyway. Note that worse seeing
again worsens the detectability of objects, since at some value, it
will distribute too much flux outwards and cause the core of the
profile to drop beyond the threshold. For a full discussion, see
Section~\ref{s:seeing}.

From these plots it becomes apparent, that the theoretical predictions
provide a good estimate of the completeness function (as the
detectable area drops below the required value at the same apparent
magnitude at which the completeness function drops off), but only
provided that seeing is taken into account.  

In the case of the high-redshift ($z=1.25$) galaxies, the detection
probability never reaches one. This results from the fact that even
for the brightest simulated objects -- with an absolute magnitude $3$
mag above $M^*$, but lower mean surface brightness -- the object's
core lies barely above the detection threshold for all magnitudes.
Therefore, noise easily scatters objects below the threshold and the
detection probability is always smaller than unity.

Looking at the predictions of the visibility function for the
exponential profiles at high redshifts, the low detection probability
surprises. Although the detectable area is much larger than the
limiting one, the completeness fraction remains low. Due to its rather
flat light-distribution, the exponential profile is much more
susceptible to distortions caused by noise than the steeper de
Vaucouleurs profile. These scattered noise pixels can then either
reduce the objects' size below the limiting radius, or form additional
false maxima leading to a mis-detection in the form of several
objects, not recognising the artificially created one.

\section{Lost-light fraction}
\label{s:lostlight}

\begin{figure*}
  \begin{minipage}{\textwidth}
  \includegraphics[width=8.4cm]{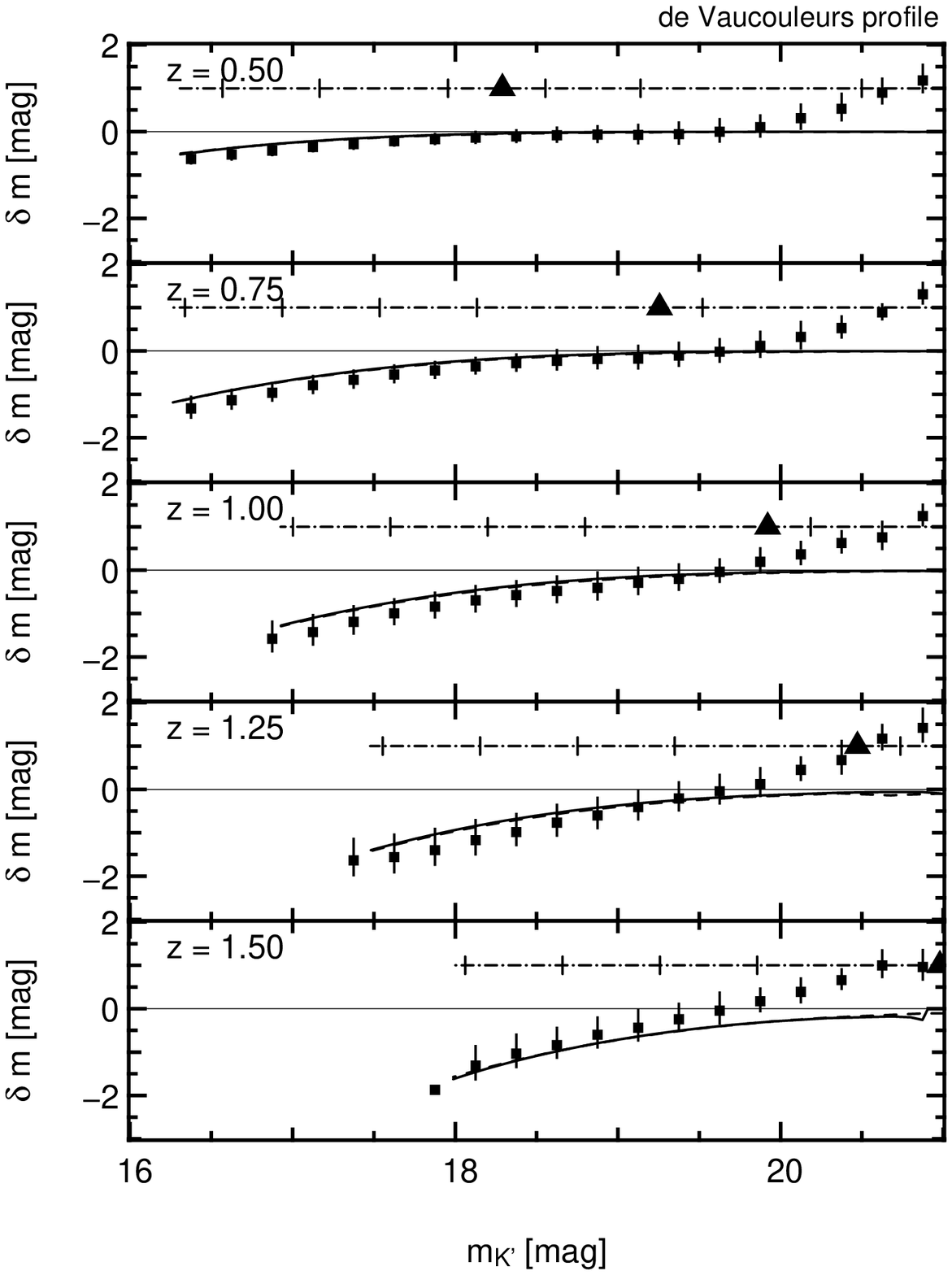}\hfill
  \includegraphics[width=8.4cm]{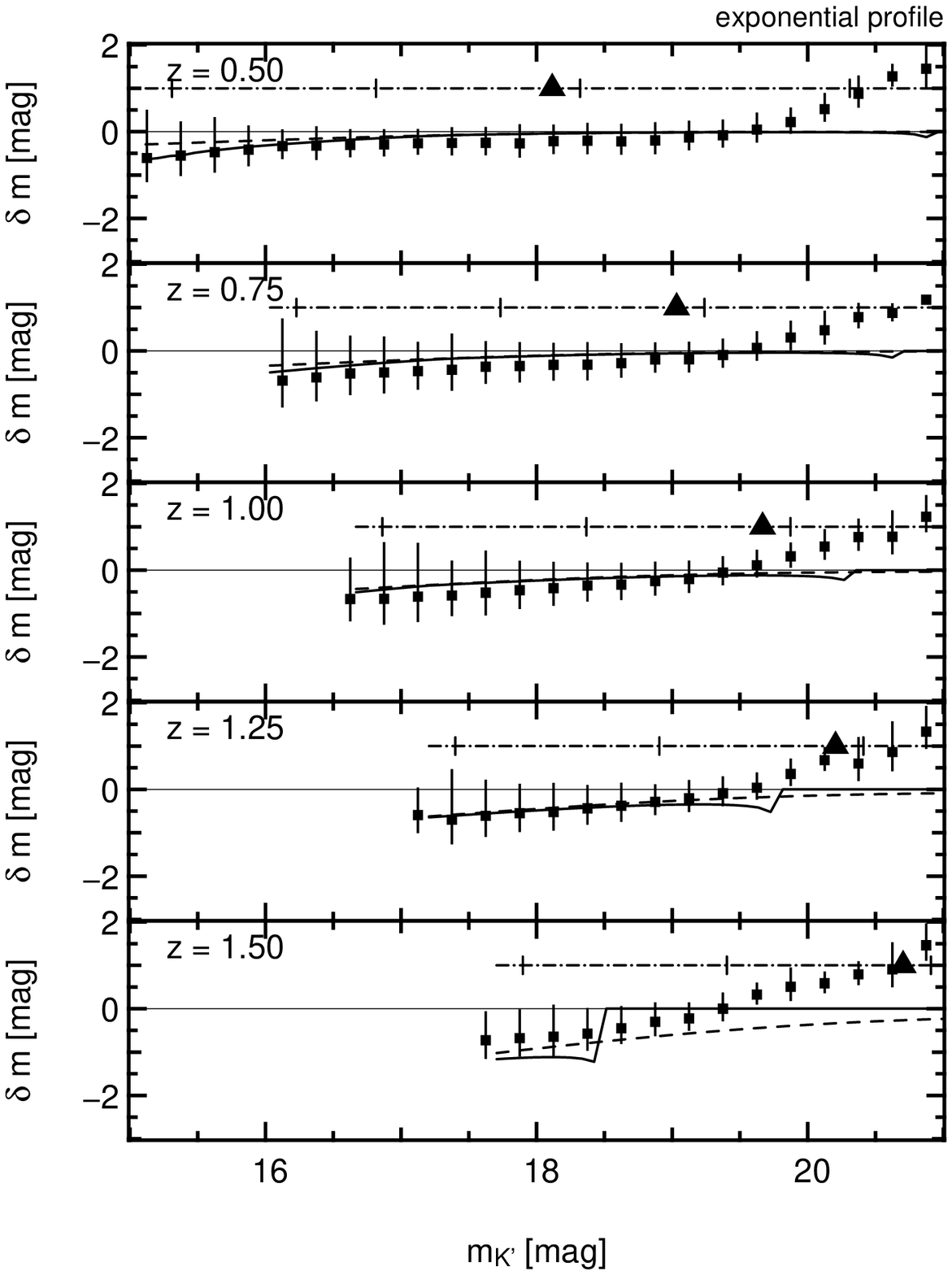}
  \end{minipage}
  \caption{Deviation of the measured magnitude from the assigned input 
    magnitude $\delta m$ for de Vaucouleurs profiles (left figure) and
    exponential profiles (right figure) with a seeing of 0.8 arcsec,
    as a function of the assigned $K'$-band input magnitude. To
    provide a clue for the width of the distribution, the rms is shown
    as errorbars. The solid line shows the theoretical prediction of
    the lost-light fraction in absence of seeing, the dashed line for
    profiles convolved with a Gaussian seeing. The horizontal scale
    (dashed-dotted line) at $\delta m = 1$ in each panel gives the effective
    radii of the simulated objects for $R_e \in \lbrace 10,
    5,2,1,0.5,0.1 \rbrace$ kpc for the de Vaucouleurs profiles at $z=0.5$,
    $r_e \in \lbrace 40,20,10,5,1,0.1 \rbrace$ kpc for the other
    elliptical galaxies and $r_e \in \lbrace 20,10,5,2,1 \rbrace$ kpc for
    the exponential profiles. The large filled triangle denotes the
    apparent magnitude and radius of an $M^*$ galaxy.}
  \label{fig:llf-20}
\end{figure*}

\begin{figure*}
  \begin{minipage}{\textwidth}
  \includegraphics[width=8.4cm]{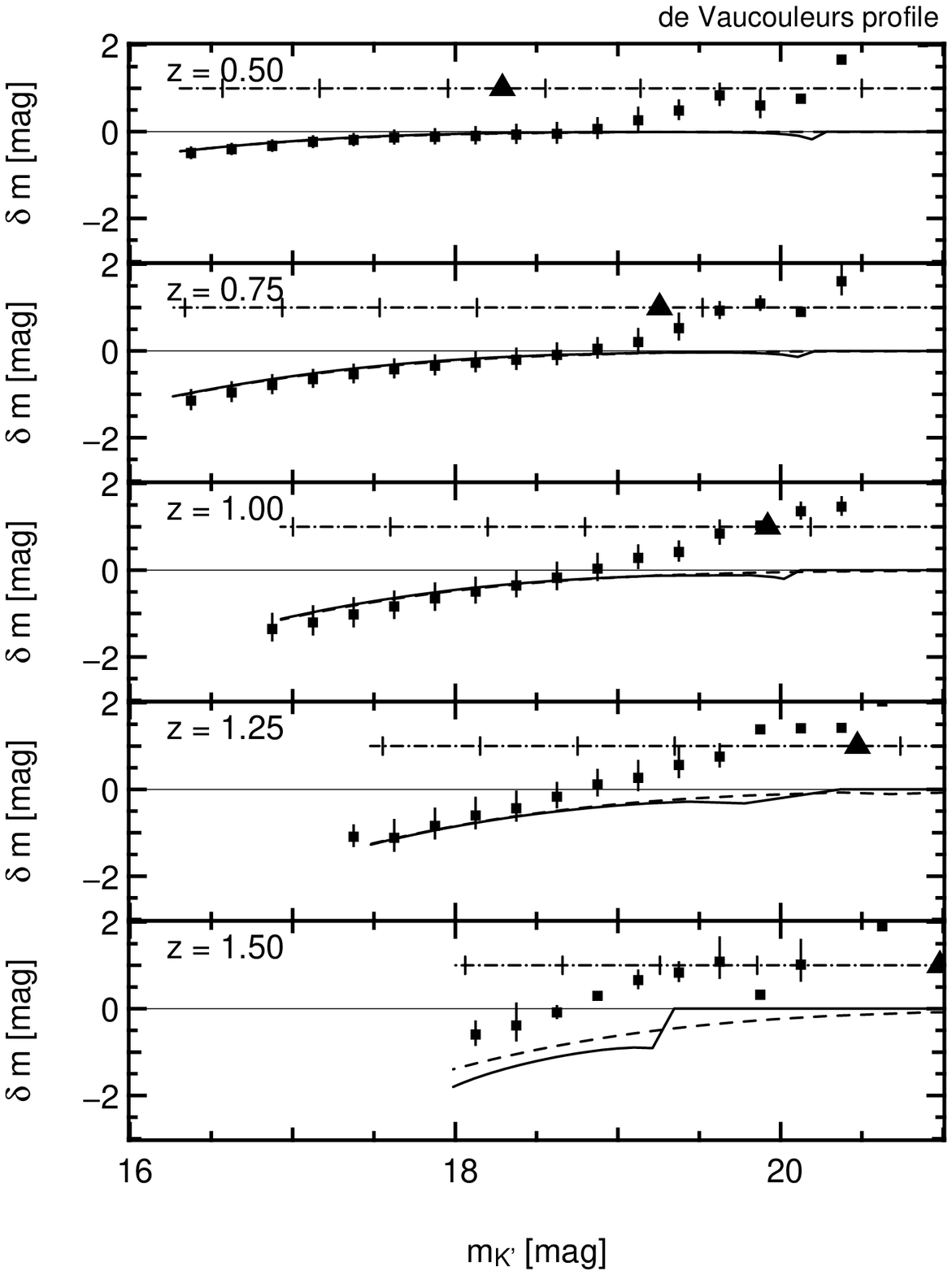}\hfill
  \includegraphics[width=8.4cm]{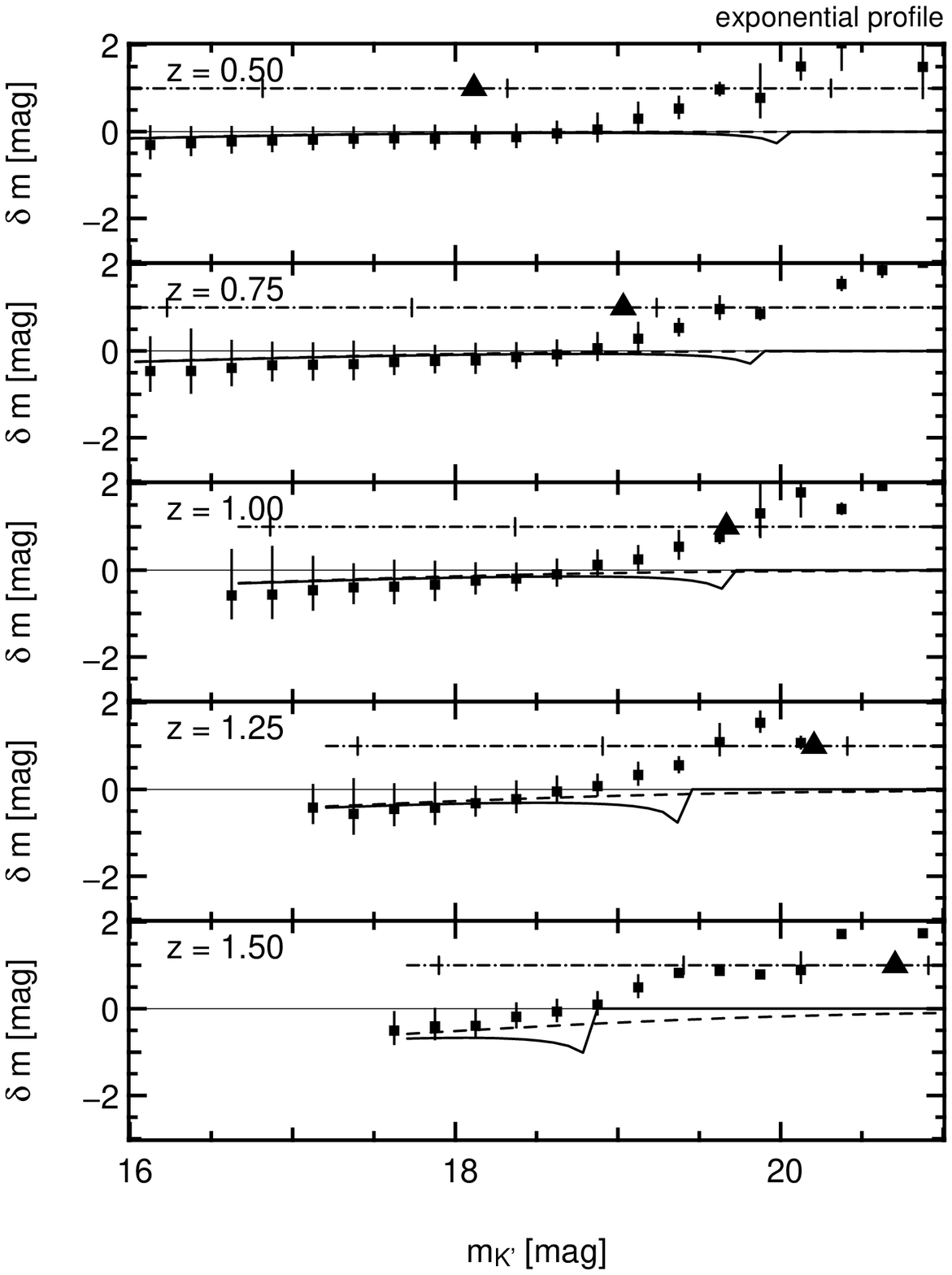}
  \end{minipage}
  \caption{Deviation of the measured magnitude from the assigned input 
    magnitude $\delta m$ for de Vaucouleurs profiles (left figure) and
    exponential profiles (right figure) with a seeing of 1.6 arcsec.
    To provide a clue for the width of the distribution, the rms is
    shown as errorbars. The solid line shows the theoretical
    prediction of the lost-light fraction in absence of seeing, the
    dashed line for profiles convolved with a Gaussian seeing.The
    horizontal scale (dash-dotted line) at $\delta m = 1$ in each panel
    gives the effective radii of the simulated objects for $R_e \in
    \lbrace 10, 5,2,1,0.5,0.1 \rbrace$ kpc for the de Vaucouleurs
    profiles at $z=0.5$, $r_e \in \lbrace 40,20,10,5,1,0.1 \rbrace$
    kpc for the other elliptical galaxies and $r_e \in \lbrace
    20,10,5,2,1 \rbrace$ kpc for the exponential profiles. The large
    filled triangle denotes the apparent magnitude and radius of an
    $M^*$ galaxy.}
  \label{fig:llf-41}
\end{figure*}

The magnitude differences between the input and the measured
magnitudes shown in Fig.~\ref{f:compl_magmag} exhibit a strong
deviation for elliptical galaxies at the bright end of the
distribution. The results from the analysis of the visibility function
can be used to calculate the lost-light fraction.  Assuming that an
object is detected out to the limiting radius $r_{lim}$ where the
surface brightness drops below the detection limit, the intensity
weighted Kron radius $r_k$ \citep{Kron80} can be calculated for
circularly symmetric objects as
\begin{equation}
  \label{eq:4}
  r_k = \frac{\sum r I(r)}{\sum I(r)}
\end{equation}

The galactic profile is then integrated out to some factor (in our
case $2.5$) times the Kron radius, and the corresponding magnitude is
calculated.  Fig.~\ref{fig:llf-20} shows the results of this
calculation of the lost-light fraction for circular symmetric de
Vaucouleurs and exponential profiles in comparison with the results
obtained from the completeness simulations.

The measured magnitudes of the elliptical galaxies deviate strongly
from the assigned input magnitudes due to underestimation of the
Kron radius caused by the steep decline of the profile in the central
detectable part. In case of the exponential profiles the differences
in the photometry are much smaller, since the flatter profile causes
the Kron-radius estimate to be closer to the true value, even if only
the inner part of the profile is detected.  

The analytically calculated lost-light fraction predicts a slightly
lower difference than the measurements show (except for the
highest-redshift bin). This can be explained by the fact, that the
theoretical approach is based on the ideal case, where the object is
detected exactly out to the maximum visible radius, and that the form
of the profile is not distorted. In reality, the area is pixelated and
integration over the pixels in calculating the Kron radius causes
smaller values. Additionally, noise causes mis-measurement of the
object size and shape.

In contrast, in the highest redshift bin ($z=1.5$) analysed here in
the context of MUNICS, the analytically calculated lost-light fraction
predicts higher deviations than actually measured in the simulated
images. These objects are only detectable due to additional pixels
being scattered above the threshold by noise (as can be seen in
Fig.~\ref{fig:vf-devauc}, which shows that the whole profile is below
the threshold).  This additional flux causes the photometry to yield
too bright magnitudes.

The same effect causes the photometry to yield too bright magnitudes
for objects at the faint end of the luminosity function in all
redshift bins. These are also too faint to be detected without the
presence of pixels scattered above the threshold by noise.

As the mean surface brightness increases towards fainter objects, and
as those are also intrinsically smaller, a larger fraction of the
total profile is visible around $M^*$, and the measurement of the
total magnitude improves.

It should be kept in mind that the largest differences occur for the
rather rare objects with absolute magnitudes 3 mag brighter then
$M^*$, while the magnitudes of $M^*$ objects are measured correctly.
In the calculation of the luminosity function, this would cause the
brightest objects to be redistributed to lower absolute magnitudes.
But again, the numbers of such bright galaxies are rather low ($\sim
10^{-4}$ times lower space density than $M^*$ objects), and thus the
contamination of the LF by these galaxies should be rather small. At
the faint end, the situation is similar. Although these objects are
much more numerous, they are actually below the detection limit.
Therefore these are only detected by chance (about 200 out of 20000
simulated objects at $z=0.5$ were detected beyond 20 mag), Thus the
influence of their wrongly measured magnitude on statistics like the
luminosity function is almost negligible.

\section{The effect of seeing}
\label{s:seeing}

\begin{figure*}
  \begin{minipage}{\textwidth}
  \includegraphics[width=8.4cm]{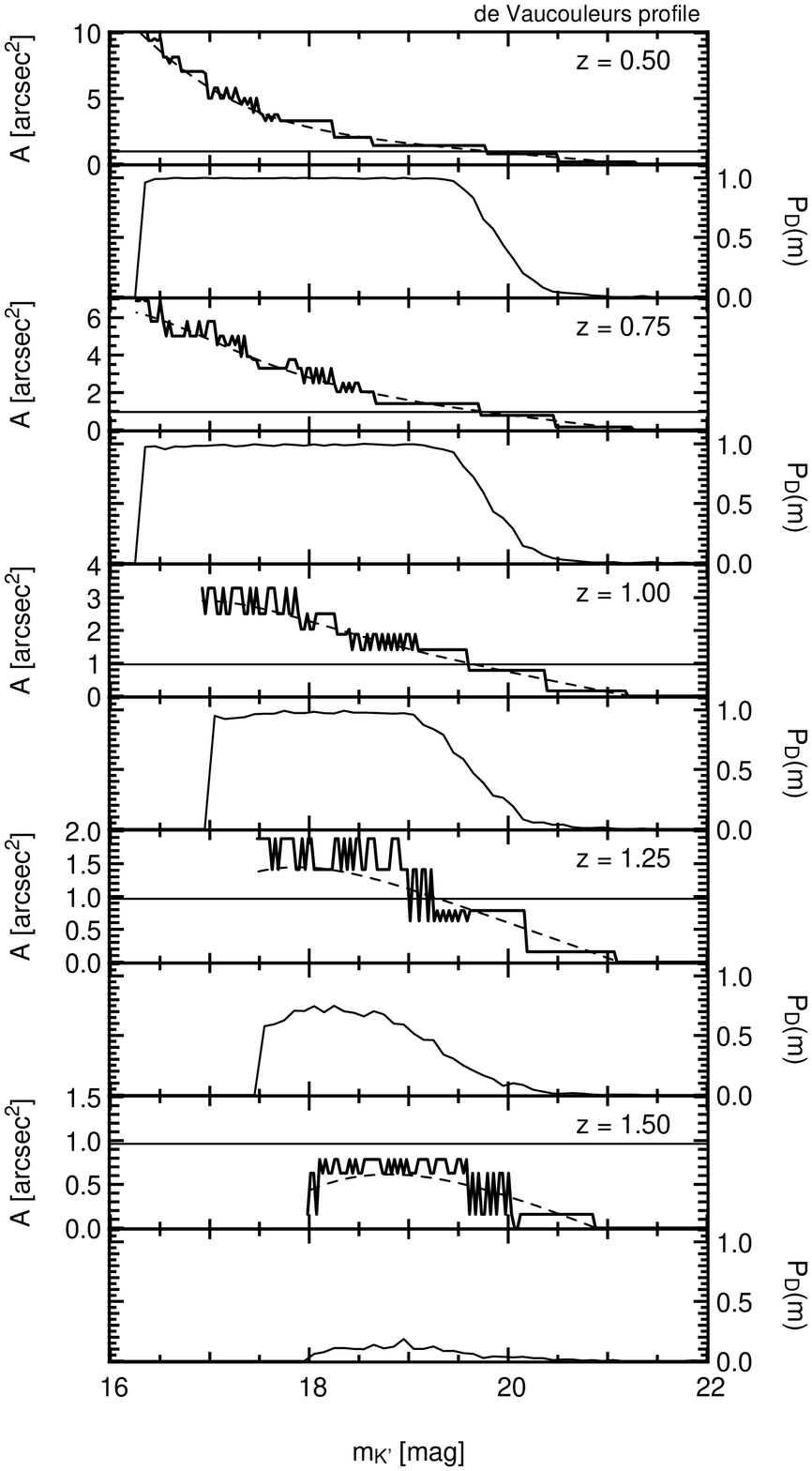}\hfill
  \includegraphics[width=8.4cm]{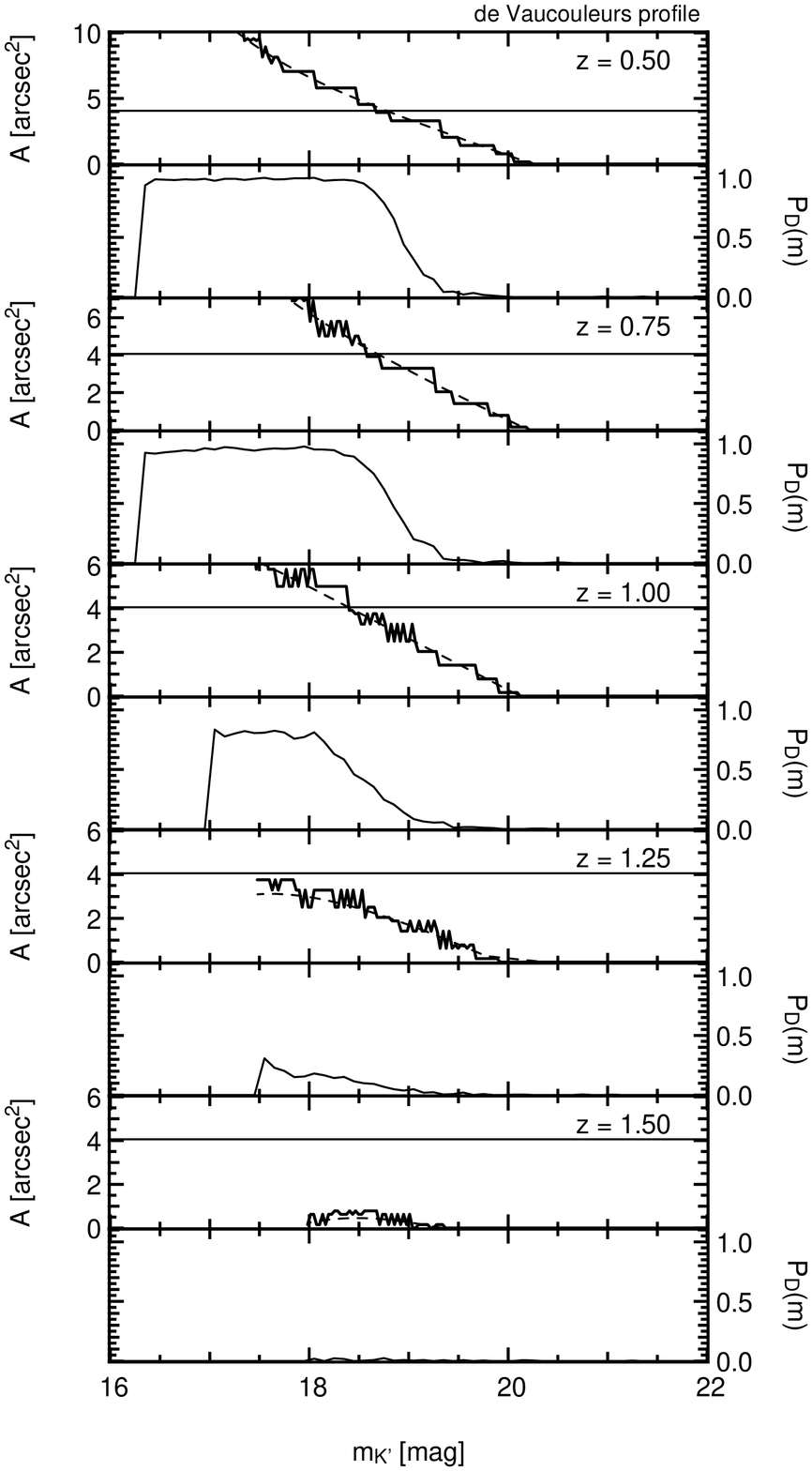}
  \end{minipage}
  \caption{Visibility function and completeness function for circularly 
    symmetric de Vaucouleurs profiles convolved with a Gaussian seeing
    of 0.8 arcsec (left figure) and 1.6 arcsec (right figure) at
    redshifts $z \in \{0.5, 0.75, 1.0, 1.25, 1.5\}$. The upper panels
    show the theoretical predictions of the visibility function in the
    analytic case (dashed) and integrated over the image pixels
    (solid). The lower panels show the object recovery fraction as a
    function of input magnitude.}
  \label{fig:vf-devauc}
\end{figure*}

\begin{figure*}
  \begin{minipage}{\textwidth}
  \includegraphics[width=8.4cm]{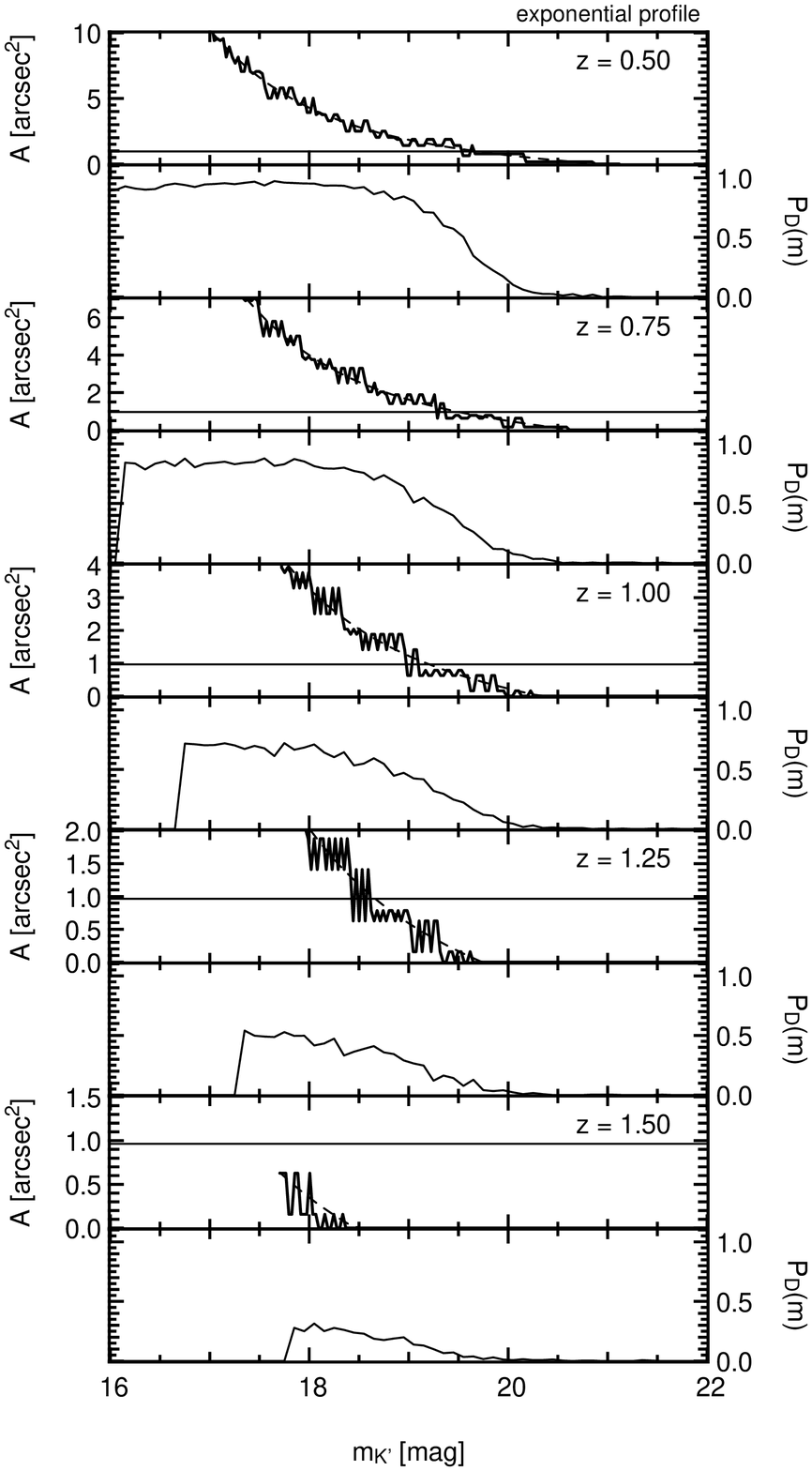}\hfill
  \includegraphics[width=8.4cm]{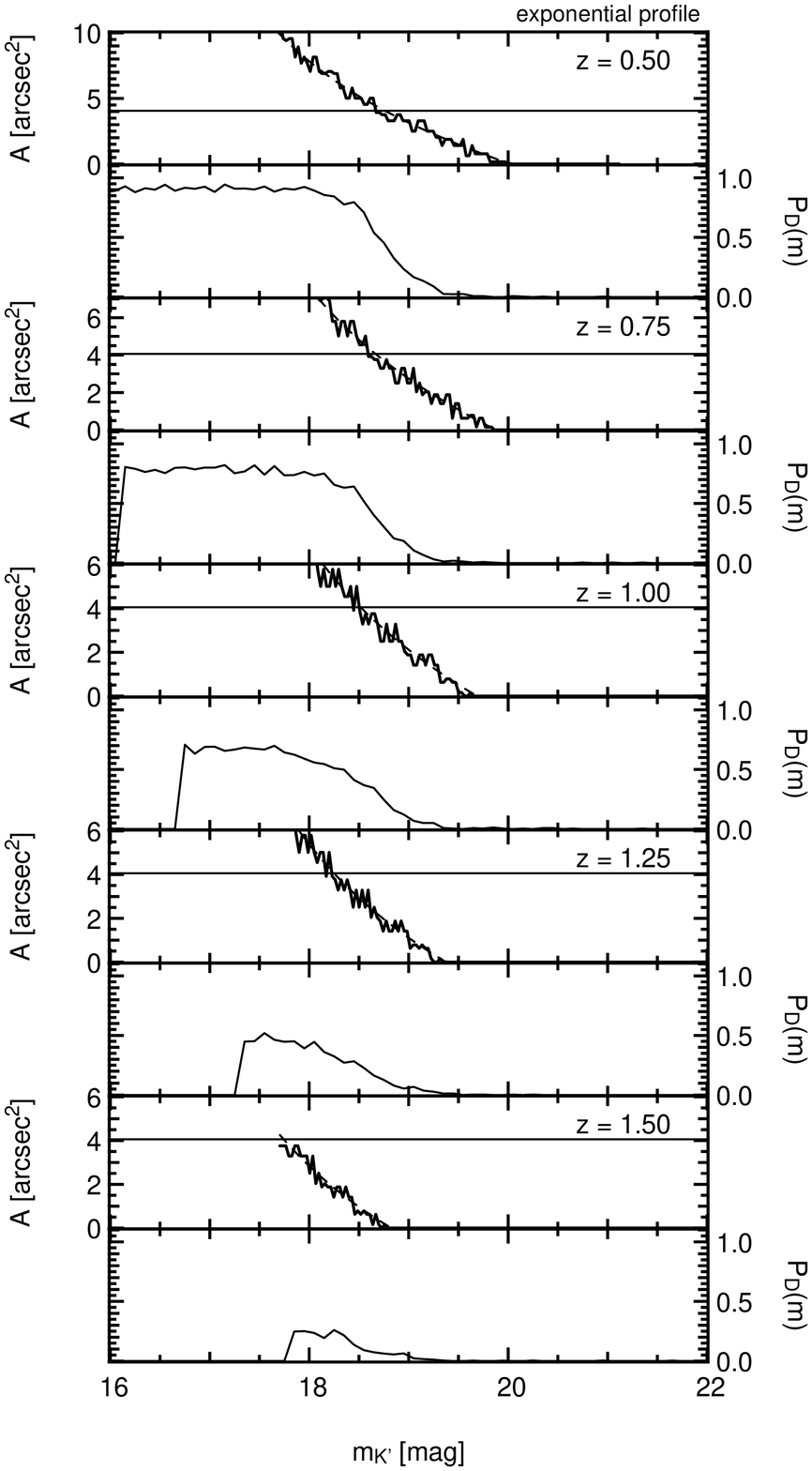}
  \end{minipage}
  \caption{Visibility function and completeness function for circularly
    symmetric exponential profiles convolved with a Gaussian seeing of
    0.8 arcsec (left figure) and 1.6 arcsec (right figure) at
    redshifts $z \in \{0.5, 0.75, 1.0, 1.25, 1.5\}$. The upper panels
    show the theoretical predictions of the visibility function in the
    analytic case (dashed) and integrated over the image pixels
    (solid). The lower panels show the object recovery fraction as a
    function of input magnitude.}
  \label{fig:vf-expdisk}
\end{figure*}

To explore the influence of seeing on the visibility, seeing convolved
galactic profiles were used to calculate the visibility function and
the lost-light fraction.  The seeing was simulated using a
two-dimensional Gaussian of the form
\begin{equation}
  \label{eq:1}
  I(\mathbf{r}) = \frac{1}{2 \pi \sigma^2} \exp \left( - \frac{\mathbf{r}^2}{2
  \sigma^2} \right) .
\end{equation}
The width of the Gaussian kernel $\sigma$ was calculated from the width
$s$ of the measured seeing PSF as
\begin{equation}
  \label{eq:2}
  \sigma = \frac{s}{\sqrt{8 \ln 2}}.
\end{equation}
The galactic profiles $I_g$ were convolved with the Gaussian resulting
in the convolved profile $I'_g$
\begin{equation}
  \label{eq:3}
  I'_g(r) = \int\limits_0^\infty \int\limits_0^{2\pi} r' I_g(r) 
  e^{ - \frac{ r^2 + {r'} ^2 - 2 r r' \cos \varphi }{2
  \sigma^2}} d\varphi dr'
\end{equation}

Using convolved profiles, the calculation of the visibility function
and the lost-light fraction were repeated on images with two different
seeing values of 0.8 and 1.6 arcsec.

Fig.~\ref{fig:vf-devauc} shows the results of the completeness
simulations and the calculation of the visibility function for de
Vaucouleurs profiles with seeing of 0.8 and 1.6 arcsec,
Fig.~\ref{fig:vf-expdisk} the same for exponential profiles.

An increase of the seeing distributes more light outward from the
central parts of the profile within the seeing disk, resulting in a
smoothed light distribution in the centre, as shown in
Fig.~\ref{f:prof}. As discussed above, moderate seeing improves
detectability of faint objects by distributing light from the bright
centre more evenly across a larger area without reducing the light in
the centre below the threshold.  As the seeing gets larger, this
effect is counter balanced by the fact that now the amount of light
redistributed away from the centre becomes so large that the central
parts of the profile falls below the detection threshold.

Fig.~\ref{fig:vf-devauc} shows this effect for the case of the steep
de Vaucouleurs profile The detection limit is fainter by roughly one
magnitude in the presence of a seeing of 0.8 arcsec compared to the
seeing-free case. On the other hand, once the seeing is as large as
1.6 arcsec, the detection limit has dropped close to the no seeing
case. In the highest-redshift bins, detectability of elliptical
galaxies is depressed below the no-seeing case, since now the
seeing-convolved profile is entirely below the detection threshold.

The exponential profiles -- shown in Fig.~\ref{fig:vf-expdisk} --
suffer from the same effects, resulting in decrease of the
completeness limit with increased seeing. The low overall recovery
fraction of exponential profiles compared to de Vaucouleurs profiles
at high redshift is caused by the flatter light distribution and the
lack of a central peak compared with the de Vaucouleurs profiles. Due
to this, the objects' appearance is much more irregular in the images
due to the significant fraction of flux in poissonian noise.  This
results, on the one hand, in pixels dropping below the detection
limit, and, on the other hand, false maxima, resulting in the
detection of two or more (distinct) structures.  These effects have
the strongest impact for the circularly symmetric (face-on) objects used
here. In the more realistic approach using randomly distributed
ellipticities -- as shown in Fig.~\ref{f:compl_results} -- the impact
of these effects is much less significant, due to the steeper apparent
profiles for inclined disks.

The lost-light fraction -- shown in Figs.~\ref{fig:llf-20}
and~\ref{fig:llf-41} -- for objects with seeing, in contrast, shows no
significant change with the increase of the seeing except close to
the completeness limit and at high redshift.

\section{$K'$-band completeness limits for the MUNICS survey}
\label{sec:values}

Here we present the results of the completeness simulations for the
fields of the MUNICS survey (see \citealp{MUNICS1}) in the $K'$-band. To
parameterise the shape of the completeness curves a combination of two
power-laws with the functional form
\begin{equation}
  \label{eq:comp_func}
  P_D(m) = \frac{p_0}{\left(\frac{m}{m_0}\right)^\alpha
    + \left(\frac{m}{m_0}\right)^\beta}
\end{equation}
was fitted to the results. The best-fitting parameters $p_0$, $m_0$,
$\alpha$ and $\beta$ are given in Table~\ref{tab:moffat_parm} for
point-like sources, Tables~\ref{tab:devauc_parm_lowz}
and~\ref{tab:devauc_parm_highz} for de Vaucouleurs profiles and
Tables~\ref{tab:expdisk_parm_lowz} and~\ref{tab:expdisk_parm_highz}
for exponential profiles. In case of the latter ones the parameters
are given separately for the five analysed redshifts.

\begin{table}
  \centering
  \begin{tabular}{l|cccc}
    Field & $p_0$ & $m_0$ & $\alpha$ & $\beta$ \\\hline
    S2F1 & 0.98 & 18.82 & 0.11 & 127.16 \\ 
    S2F5 & 0.98 & 19.12 & 0.08 & 143.12 \\ 
    S3F5 & 0.96 & 19.25 & 0.16 & 140.07 \\ 
    S4F1 & 0.99 & 19.08 & 0.01 & 101.10 \\ 
    S5F1 & 0.99 & 19.09 & 0.05 & 144.70 \\ 
    S5F5 & 0.99 & 19.11 & 0.04 & 148.81 \\ 
    S6F1 & 0.99 & 18.75 & 0.00 & 149.33 \\ 
    S6F5 & 0.97 & 19.25 & 0.12 & 142.99 \\ 
    S7F5 & 0.98 & 19.06 & 0.06 & 92.30 \\ 
  \end{tabular}
  \caption{Parameters of the best fit of equation~(\ref{eq:comp_func})
    to the results obtained from the completeness simulations for
    point-like sources. The normalisation $p_0$ provides a rough
    estimate of the height of the plateau the detection probability
    reaches. The break magnitude $m_0$ can be used as an approximation
    of the $50\%$ completeness limit, if a higher maximum detection
    rate is reached.}
  \label{tab:moffat_parm}
\end{table}

\begin{table*}
  \centering
  \begin{tabular}{l|cccccccc}
    Field & \multicolumn{4}{c}{$z=0.5$} & \multicolumn{4}{c}{$z=0.75$} \\
    & $p_0$ & $m_0$ & $\alpha$ & $\beta$ & $p_0$ & $m_0$ & $\alpha$ &
    $\beta$ \\\hline
    S2F1 & 
    1.02 & 19.22 & -0.32 & 110.50 & 
    1.01 & 19.12 & -0.31 & 102.47 \\
    S2F5 & 
    1.01 & 19.54 & -0.15 & 118.58 & 
    1.01 & 19.44 & -0.20 & 114.66 \\
    S3F5 & 
    1.01 & 19.65 & -0.20 & 139.21 & 
    1.02 & 19.60 & -0.28 & 111.96 \\ 
    S4F1 & 
    1.00 & 19.53 & -0.04 & 98.09 & 
    0.99 & 19.42 & -0.09 & 82.18 \\
    S5F1 & 
    1.01 & 19.48 & -0.18 & 129.05 & 
    1.02 & 19.40 & -0.38 & 110.91 \\ 
    S5F5 & 
    1.01 & 19.52 & -0.16 & 128.85 & 
    1.02 & 19.44 & -0.40 & 107.08 \\
    S6F1 & 
    1.01 & 19.14 & -0.13 & 127.56 & 
    1.00 & 19.03 & -0.34 & 126.05 \\
    S6F5 & 
    1.01 & 19.67 & -0.11 & 127.00 & 
    1.01 & 19.61 & -0.21 & 121.12 \\
    S7F5 & 
    1.00 & 19.48 & -0.08 & 87.53 & 
    1.00 & 19.40 & -0.22 & 74.91 \\
  \end{tabular}
  \caption{Parameters of the fit of equation~(\ref{eq:comp_func})
    against the results obtained from the completeness simulations for
    de Vaucouleurs profiles at redshifts $z = 0.5$ and $z = 0.75$.
    Shown are the best-fit parameters of the formula given in
    equation~(\ref{eq:comp_func}) $p_0,m_0,\alpha$, and $\beta$. The
    normalisation $p_0$ provides a rough estimate of the height of the
    plateau the detection probability reaches. The break magnitude
    $m_0$ can be used as an approximation of the $50\%$ completeness
    limit, if a higher maximum detection rate is reached.}
  \label{tab:devauc_parm_lowz}
  \begin{tabular}{l|cccccccccccc}
    Field & \multicolumn{4}{c}{$z=1.0$} & \multicolumn{4}{c}{$z=1.25$} &
    \multicolumn{4}{c}{$z=1.5$} \\
    & $p_0$ & $m_0$ & $\alpha$ & $\beta$ & $p_0$ & $m_0$ &
    $\alpha$ & $\beta$ & $p_0$ & $m_0$ & $\alpha$ & $\beta$ \\ \hline
    S2F1 & 
    0.99 & 18.93 & -1.22 & 76.05 & 
    0.34 & 18.93 & 5.92 & 85.92 & 
    0.06 & 19.02 & -6.91 & 56.93 \\ 
    S2F5 & 
    1.02 & 19.29 & -0.70 & 97.96 & 
    0.80 & 18.98 & -2.44 & 70.49 & 
    0.14 & 19.18 & -1.45 & 64.29 \\
    S3F5 & 
    1.01 & 19.45 & -0.55 & 106.26 & 
    0.95 & 19.14 & -2.49 & 70.63 & 
    0.27 & 19.12 & -4.79 & 59.58 \\
    S4F1 & 
    0.86 & 19.30 & 0.38 & 77.23 & 
    0.59 & 18.98 & -3.16 & 52.67 & 
    0.12 & 18.91 & -9.17 & 41.61 \\
    S5F1 & 
    1.01 & 19.24 & -0.72 & 94.38 & 
    0.80 & 18.92 & -2.44 & 65.87 & 
    0.26 & 18.21 & -41.32 & 33.96 \\
    S5F5 & 
    1.02 & 19.29 & -0.83 & 90.94 & 
    0.86 & 18.95 & -3.57 & 60.75 & 
    0.05 & 19.59 & 13.68 & 113.45 \\
    S6F1 & 
    0.96 & 18.77 & -0.99 & 74.39 & 
    0.19 & 18.83 & 8.28 & 65.91 & 
    -- & -- & -- & -- \\
    S6F5 & 
    1.01 & 19.50 & -0.44 & 107.66 & 
    0.96 & 19.22 & -1.27 & 68.88 & 
    0.36 & 18.99 & -1.59 & 50.96 \\
    S7F5 & 
    0.91 & 19.31 & 0.02 & 71.80 & 
    0.78 & 19.06 & -3.14 & 58.76 & 
    0.13 & 19.52 & 6.32 & 78.10 \\
  \end{tabular}
  \caption{Parameters of the fit of equation~(\ref{eq:comp_func})
    against the results obtained from the completeness simulations for
    de Vaucouleurs profiles at three distinct redshifts $z \in \lbrace
    1.0, 1.25, 1.50 \rbrace$. Shown are the best-fit parameters of
    the formula given in equation~(\ref{eq:comp_func}) $p_0,m_0,\alpha$,
    and $\beta$. The normalisation $p_0$ provides a rough estimate of
    the height of the plateau the detection probability reaches. The
    break magnitude $m_0$ can be used as an approximation of the
    $50\%$ completeness limit, if a higher maximum detection rate is
    reached. The detection probability in field S6F1 at redshift
    $z=1.50$ is too low to allow a reliable fit to the data.}
  \label{tab:devauc_parm_highz}
\end{table*}

\begin{table*}
  \centering
  \begin{tabular}{l|cccccccc}
    Field & \multicolumn{4}{c}{$z=0.5$} & \multicolumn{4}{c}{$z=0.75$} \\
    & $p_0$ & $m_0$ & $\alpha$ & $\beta$ & $p_0$ & $m_0$ & $\alpha$ &
    $\beta$ \\\hline
     S2F1 & 
     0.97 & 19.12 & -0.07 & 103.55 & 
     0.92 & 19.05 & 0.07 & 90.73 \\
     S2F5 & 
     0.98 & 19.39 & -0.05 & 111.26 & 
     0.95 & 19.33 & -0.04 & 100.28 \\
     S3F5 & 
     0.99 & 19.54 & -0.05 & 105.69 & 
     0.97 & 19.44 & -0.11 & 98.81 \\
     S4F1 & 
     0.95 & 19.34 & 0.05 & 81.52 & 
     0.89 & 19.22 & 0.10 & 75.30 \\
     S5F1 & 
     0.99 & 19.35 & -0.05 & 101.58 & 
     0.96 & 19.26 & -0.16 & 90.01 \\ 
     S5F5 & 
     0.99 & 19.38 & -0.11 & 107.76 & 
     0.94 & 19.31 & 0.00  & 95.69 \\
     S6F1 & 
     0.98 & 18.98 & -0.21 & 104.80 & 
     0.90 & 18.93 & -0.07 & 86.86 \\ 
     S6F5 & 
     0.99 & 19.56 & 0.00 & 113.58 & 
     0.96 & 19.48 & -0.03 & 99.99 \\
     S7F5 & 
     0.98 & 19.37 & -0.05 & 80.93 & 
     0.95 & 19.27 & -0.11 & 67.06 \\
  \end{tabular}
  \caption{Parameters of the fit of equation~(\ref{eq:comp_func})
    against the results obtained from the completeness simulations for
    exponential profiles at at redshifts $z = 0.5$ and $z = 0.75$.
    Shown are the best-fit parameters of
    the formula given in equation~(\ref{eq:comp_func}) $p_0,m_0,\alpha$,
    and $\beta$. The normalisation $p_0$ provides a rough estimate of
    the height of the plateau the detection probability reaches. The
    break magnitude $m_0$ can be used as an approximation of the
    $50\%$ completeness limit, if a higher maximum detection rate is reached.}
  \label{tab:expdisk_parm_lowz}

  \begin{tabular}{l|cccccccccccc}
    Field & \multicolumn{4}{c}{$z=1.0$} & \multicolumn{4}{c}{$z=1.25$} &
    \multicolumn{4}{c}{$z=1.5$} \\
    & $p_0$ & $m_0$ & $\alpha$ & $\beta$ & $p_0$ & $m_0$ &
    $\alpha$ & $\beta$ & $p_0$ & $m_0$ & $\alpha$ & $\beta$ \\ \hline
     S2F1 & 
     0.84 & 19.00 & 0.32 & 89.53 & 
     0.69 & 18.98 & 1.38 & 88.07 & 
     0.55 & 18.92 & 3.27 & 83.59 \\
     S2F5 & 
     0.86 & 19.28 & 0.41 & 89.29 & 
     0.80 & 19.14 & 0.24 & 73.32 & 
     0.67 & 19.08 & 0.48 & 75.83 \\
     S3F5 & 
     0.92 & 19.39 & -0.18 & 90.77 & 
     0.82 & 19.31 & 0.29 & 81.43 & 
     0.64 & 19.27 & 1.90 & 83.07 \\
     S4F1 & 
     0.84 & 19.14 & 0.08 & 61.14 & 
     0.67 & 19.08 & 1.28 & 65.21 & 
     0.47 & 19.12 & 3.43 & 72.29 \\
     S5F1 & 
     0.87 & 19.20 & 0.22 & 95.03 & 
     0.77 & 19.14 & 0.50 & 79.92 & 
     0.57 & 19.11 & 2.67 & 82.73 \\
     S5F5 & 
     0.88 & 19.22 & 0.10 & 87.80 & 
     0.77 & 19.17 & 0.31 & 81.99 & 
     0.61 & 19.13 & 1.30 & 84.03 \\
     S6F1 & 
     0.81 & 18.84 & 0.07 & 82.68 & 
     0.60 & 18.86 & 2.46 & 93.53 & 
     0.36 & 18.90 & 7.54 & 120.92 \\
     S6F5 & 
     0.93 & 19.42 & -0.07 & 90.57 & 
     0.85 & 19.37 & -0.06 & 88.12 & 
     0.68 & 19.32 & 1.25 & 93.77 \\
     S7F5 & 
     0.86 & 19.26 & 0.19 & 73.53 & 
     0.74 & 19.22 & 0.84 & 75.93 & 
     0.52 & 19.24 & 3.60 & 72.31 \\
  \end{tabular}
  \caption{Parameters of the fit of equation~(\ref{eq:comp_func})
    against the results obtained from the completeness simulations for
    exponential profiles at three distinct redshifts $z \in \lbrace
    1.0, 1.25, 1.50 \rbrace$.  Shown are the best-fit parameters of
    the formula given in equation~(\ref{eq:comp_func}) $p_0,m_0,\alpha$,
    and $\beta$. The normalisation $p_0$ provides a rough estimate of
    the height of the plateau the detection probability reaches. The
    break magnitude $m_0$ can be used as an approximation of the
    $50\%$ completeness limit, if a higher maximum detection rate is reached.}
  \label{tab:expdisk_parm_highz}
\end{table*}

\section{Summary}

We presented the results of extensive completeness simulations for
imaging surveys of faint galaxies, taking into account their known size --
surface-brightness relations. The absolute magnitudes of the
elliptical galaxies simulated as de Vaucouleurs profiles following a
local fundamental-plane relation. The disk galaxies represented by
pure exponential profiles were modelled according to a Freeman law.
These physical parameters were converted into apparent sizes,
depending on the simulated redshift of the object. The simulations
were carried out for the $K'$-band images of the MUNICS survey, but
the deficiencies found should remain valid in other comparable surveys
as long as a threshold based detection algorithm and an adaptive
isophotal magnitude measurement are used.

We find a strong bias against bright $r^1/4$ profiles at higher
redshifts in the rate of detection as well in the photometry of the
objects. At low redshifts the completeness fraction is comparable to
point-like sources. The measured magnitudes underestimate the true
magnitudes at the bright end of the distribution. The sharp maximum in
the centre of the profile results in an underestimate of the objects'
radius leading to a too small isophotal radius in the magnitude
measurement. At high redshifts the detection probability does not
reach one even for the brightest objects, forming a plateau at lower
values. This stems from the correlation of surface brightness and
effective radius along the fundamental-plane relation.  Toward the
bright end the object's surface brightness decreases, although the
total magnitude increases.

The detection of spiral galaxies at high redshift shows similar
deficiencies. Here the effective radius increases with magnitude, but
the intrinsic mean surface brightness stays constant as predicted by
the Freeman law. The rather flat light distribution of the
exponential profile helps to estimate the object's radius correctly
and accordingly to measure the magnitudes without large errors.

These results are in general agreement with the prediction of the
visibility function theory~\citep{DP1983} and the brightness defects
predicted by~\citet{Dalcanton1998} -- except in the two most distant
redshift-bins, where the predictions break down -- using calculations
of the lost-light fraction, as long as the effects of seeing are taken
into account.

With equations~(\ref{eq:get_compl}) and~(\ref{eq:get_compl2}) we provide a
rather simple way to estimate the $\sim 50 \%$ completeness limit of
point-like sources represented by Moffat profiles.

\section*{Acknowledgements}

The authors would like to thank the staff at Calar Alto Observatory
and McDonald Observatory for their extensive support during the many
observing runs of this project. The authors also thank the anonymous
referee for his comments, which helped to improve the presentation of
the paper. This research has made use of NASA's
Astrophysics Data System (ADS) Abstract Service and the NASA/IPAC
Extragalactic Database (NED).  The MUNICS project was supported by the
Deutsche Forschungsgemeinschaft, \textit{Sonderforschungsbereich 375,
  Astroteilchenphysik}.

\bibliography{mnrasmnemonic,literature} \bibliographystyle{mn2e}

\label{lastpage}

\end{document}